\documentclass[amsmath,amssymb,aps,floats,prx,twocolumn,showpacs,superscriptaddress,longbibliography]{revtex4-1}

\usepackage{graphicx}
\usepackage{amssymb}
\usepackage{epstopdf}
\usepackage{color}

\DeclareGraphicsExtensions{.pdf, .png, .jpg, .eps}

\begin{document}

\title{Ultrafast Frustration-Breaking and Magnetophononic Driving of  
Singlet Excitations in a Quantum Magnet}

\author{F. Giorgianni}
\affiliation{Paul Scherrer Institute, CH-5232 Villigen-PSI, Switzerland}
\affiliation{Institute of Applied Physics, University of Bern, CH-3012 Bern, 
Switzerland}
\author{B. Wehinger}
\affiliation{Paul Scherrer Institute, CH-5232 Villigen-PSI, Switzerland}
\affiliation{Department of Quantum Matter Physics, University of Geneva, 
CH-1211 Geneva 4, Switzerland}
\affiliation{European Synchrotron Radiation Facility, 71 Av. des Martyrs, 
38000 Grenoble, France}
\author{S. Allenspach} 
\affiliation{Paul Scherrer Institute, CH-5232 Villigen-PSI, Switzerland}
\affiliation{Department of Quantum Matter Physics, University of Geneva, 
CH-1211 Geneva 4, Switzerland}
\author{N. Colonna}
\affiliation{Paul Scherrer Institute, CH-5232 Villigen-PSI, Switzerland}
\affiliation{National Centre for Computational Design and Discovery of Novel 
Materials (MARVEL), Ecole Polytechnique F\'ed\'erale de Lausanne (EPFL), 
CH-1015 Lausanne, Switzerland}
\author{C. Vicario} 
\affiliation{Paul Scherrer Institute, CH-5232 Villigen-PSI, Switzerland}
\author{P. Puphal} 
\affiliation{Paul Scherrer Institute, CH-5232 Villigen-PSI, Switzerland}
\affiliation{Max Planck Institute for Solid State Research, Heisenbergstrasse 
1, 70569 Stuttgart, Germany}
\author{E. Pomjakushina} 
\affiliation{Paul Scherrer Institute, CH-5232 Villigen-PSI, Switzerland}
\author{B. Normand}
\affiliation{Paul Scherrer Institute, CH-5232 Villigen-PSI, Switzerland}
\affiliation{Lehrstuhl f\"ur Theoretische Physik I, Technische Universit\"at 
Dortmund, Otto-Hahn-Strasse 4, 44221 Dortmund, Germany}
\affiliation{Institute of Physics, Ecole Polytechnique F\'ed\'erale de 
Lausanne (EPFL), CH-1015 Lausanne, Switzerland}
\author{Ch. R\"uegg}
\affiliation{Paul Scherrer Institute, CH-5232 Villigen-PSI, Switzerland}
\affiliation{Department of Quantum Matter Physics, University of Geneva, 
CH-1211 Geneva 4, Switzerland}
\affiliation{Institute of Physics, Ecole Polytechnique F\'ed\'erale de 
Lausanne (EPFL), CH-1015 Lausanne, Switzerland}
\affiliation{Institute of Quantum Electronics, ETH Z\"urich, CH-8093 
H\"onggerberg, Switzerland}

\begin{abstract}

Ideal magnetic frustration forms the basis for the emergence of exotic 
quantum spin states that are entirely nonmagnetic. Such singlet spin states 
are the defining feature of the Shastry-Sutherland model, and of its faithful 
materials realization in the quantum antiferromagnet SrCu$_2$(BO$_3$)$_2$. 
To address these states on ultrafast timescales, despite their lack of any 
microscopic order parameter, we introduce a nonlinear magnetophononic 
mechanism to alter the quantum spin dynamics by driving multiple optical 
phonon modes coherently and simultaneously. We apply intense terahertz pulses 
to create a nonequilibrium modulation of the magnetic interactions that breaks 
the ideal frustration of SrCu$_2$(BO$_3$)$_2$, such that previously forbidden 
physics can be driven in a coherent manner. Specifically, this driving 
populates a purely magnetic excitation, the singlet branch of the two-triplon 
bound state, by resonance with the difference frequency of two pumped phonons. 
Our results demonstrate how light-driven phonons can be used for the ultrafast 
and selective manipulation of interactions in condensed matter, even at 
frequencies far from those of the pump spectrum, offering valuable additional  
capabilities for the dynamical control of quantum many-body phenomena. 
\end{abstract}

\maketitle

\section{Introduction}
\label{si}

Using ultrafast lasers to access all the intrinsic interaction timescales 
of correlated quantum materials opens a new window on fundamental processes in 
nonequilibrium many-body physics. Coherent light sources developed to combine 
ultrafast time structure and high intensity at the appropriate terahertz (THz) 
or infrared (IR) frequencies \cite{Salen19,Nicoletti16} have been used in 
complex condensed matter to enhance superconductivity \cite{Mitrano16}, drive 
metal-insulator transitions \cite{Caviglia12}, manipulate multiferroic order 
\cite{Kubacka14}, and ``Floquet engineer'' the electronic band structure 
\cite{Oka19}. This type of dynamical driving, by which different and 
unconventional static and dynamic properties are conferred on a quantum 
state, has the potential to reveal many previously hidden or unknown phenomena
\cite{Basov17,tkcgis21}.

In ultrafast magnetism, the magnetic field of a light pulse can drive 
precessional spin dynamics and spin waves in ordered antiferromagnets 
\cite{Vicario13,Kampfrath11}, while the electric field can modify the magnetic 
interactions \cite{Mikhaylovskiy15}. Strong lattice excitations have been used 
to melt magnetic order \cite{Foerst15} and to induce spin waves through an 
effective magnetic field \cite{Nova17}. The concept of magnetophononics, 
defined as the resonant modulation of magnetic (super)exchange interactions 
by ultrafast coherent lattice displacements, has to date been discussed only 
in theory \cite{Fechner18}. Although phononic effects observed in ordered 
magnetic materials have been ascribed in part to exchange interactions 
\cite{Afanasiev21} or fully to crystal-field effects \cite{Disa20}, they 
remain some orders of magnitude slower than the driving phonons. While the 
ultrafast manipulation of magnetically ordered phases is developing towards 
applications in spintronics \cite{kkr10,Nemec18}, the situation in quantum 
magnets that lack any magnetic order remains largely unexplored. 

\begin{figure}[t]
\includegraphics[width=\columnwidth]{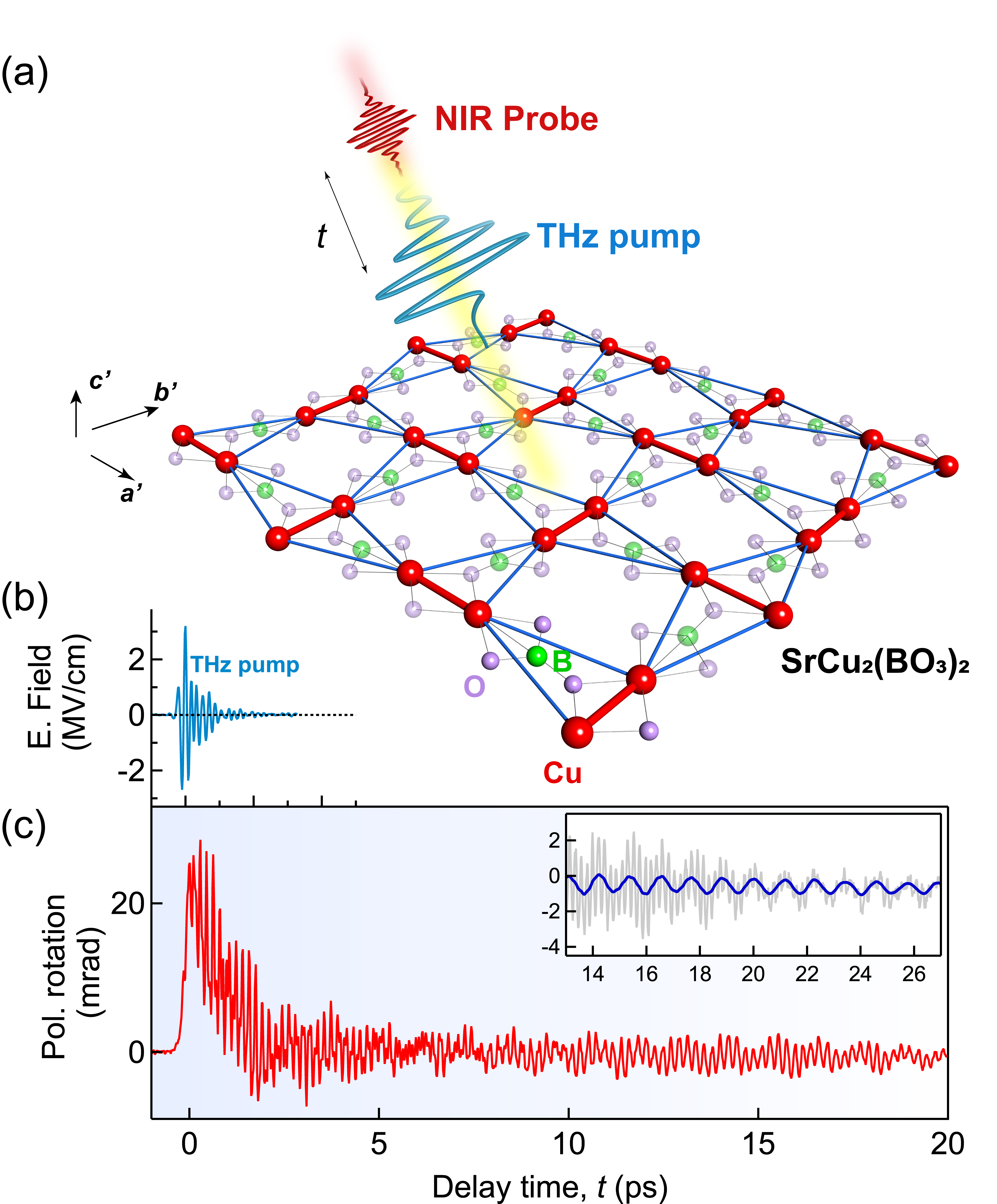}
\caption{{\bf Coherent lattice control in the time domain.} (a) An intense 
THz pulse (blue) with electric field polarized linearly along the $a'$ axis 
of a SrCu$_2$(BO$_3$)$_2$ crystal provides coherent driving of dipole-active 
lattice vibrations. A femtosecond near-IR (NIR) probe pulse (red) measures 
the polarization changes. (b-c) THz electric field and polarization rotation 
of the probe as functions of the delay time, $t$. Inset: THz-driven dynamics 
after filtering of the fast component to reveal coherent oscillations 
associated with the low-lying magnetic excitation (dark blue).}
\label{uqmf1}
\end{figure}

In this work we begin the quest to control the properties of nonordered 
quantum magnetic materials using ultrafast coherent light pulses. The 
paradigm of ideal frustration is the fundamental ingredient in all of the 
complex quantum many-body states in magnetism, most of which emerge from 
rather simple spin Hamiltonians \cite{Sachdev08}. Its most basic form is 
geometrical frustration, which has been realized in a wide range of materials 
hosting antiferromagnetically interacting spins in the triangle-based motifs 
of the kagome, pyrochlore, Shastry-Sutherland, and other lattices. More complex 
forms of ideal frustration have been produced using magnetic interactions that 
are anisotropic in spin or real space, examples including spin ices, SU(N) 
magnets, and (proximate) Kitaev systems \cite{Savary16}. However, the 
characteristic properties of the resulting ground and excited states, which 
can include both gapped and gapless quantum spin liquids \cite{Broholm20}, 
fractional quasiparticles, topological order, and long-ranged entanglement 
\cite{Savary16}, are often undetectable by the conventional probes of 
experimental condensed matter. This makes them ideal candidates for ultrafast 
probing. 

For our study (Fig.~\ref{uqmf1}) we choose SrCu$_2$(BO$_3$)$_2$, an archetypal 
quantum magnetic material whose physics is dominated by local quantum mechanical
singlet states \cite{Miyahara03}. The singlet encapsulates the essence of 
quantum magnetism, where the fluctuating spin variables combine into both 
local and global states of especially low energy that have no external 
magnetic properties \cite{Anderson73}. The ideally frustrated geometry of 
SrCu$_2$(BO$_3$)$_2$ [Fig.~\ref{uqmf1}(a)] realizes a spin model formulated 
by Shastry and Sutherland specifically for its exact dimer-singlet ground 
state \cite{Shastry81}, and if an applied pressure is used to alter the 
interaction parameters then it undergoes a first-order quantum phase 
transition (QPT) to a four-site ``plaquette'' singlet state \cite{Corboz13,
Zayed17,Guo20,Larrea21}. This ideal magnetic frustration also causes 
SrCu$_2$(BO$_3$)$_2$ to display an anomalous spectrum of spin excitations 
and complex phase transitions both in an applied magnetic field 
\cite{Kageyama99,Takigawa13,Haravifard16} and as a function of 
temperature \cite{Larrea21}. 

For the goal of ultrafast modulation of magnetic properties in nonordered 
materials such as SrCu$_2$(BO$_3$)$_2$, the magnetophononic mechanism is an 
obvious candidate. Experiments applying static pressure to quantum magnets 
have created novel ground and excited states \cite{Zayed17,Rueegg08} and have 
controlled QPTs in both localized \cite{Merchant14} and itinerant magnetic 
systems \cite{Uhlarz04}, demonstrating not only the sensitivity of the magnetic 
interactions to the atomic positions but also the potential for qualitatively 
new dynamical phenomena. In our study it is important to stress the distinction 
between magnetophononics, which is a resonant modulation of the magnetic 
interactions by harmonic phonons \cite{Fechner18}, and ``nonlinear phononics'' 
\cite{Foerst11}, which exploits the anharmonic potential of large-amplitude 
phonons. The latter offers a route for combining phonons at their sum and 
difference frequencies, and the lattice distortions it allows have been 
applied in ordered magnets to alter their static properties \cite{Nova17}. 
The former relies on the complex dependence of the magnetic interactions on 
the periodically varying lattice coordinates to effect a dynamical coupling 
to the spin sector. We will observe both mechanisms at work in our experiments, 
but only the magnetophononic mechanism drives the spin sector, and thus our 
focus lies here. 

Nevertheless, modulating an interaction, $J$, at some available phonon 
frequency, $\omega_i$, does not constitute control of dynamical properties: 
{\it a priori} there is no match between the energy scales of the dominant 
IR-active phonon modes and of the elementary magnetic excitations in any 
material, and we will see that SrCu$_2$(BO$_3$)$_2$ is a case in point. 
To achieve such frequency matching, we extend magnetophononics to the 
nonlinear regime, where sums and differences of the available phonon 
frequencies span a wide energy range, but extremely intense electric fields 
are required. By using coherent THz pulses to drive IR-active phonons in 
SrCu$_2$(BO$_3$)$_2$ (Fig.~\ref{uqmf1}), we demonstrate experimentally how 
the leading difference frequency creates a nonequilibrium occupation of the 
lowest excited singlet state. We establish the theoretical framework for 
the origin of this phenomenon, in the breaking of ideal magnetic frustration 
within the driven lattice structure, which we verify by density functional 
theory (DFT) calculations.

The structure of this article is as follows. In Sec.~\ref{sscbo} we review the 
properties of SrCu$_2$(BO$_3$)$_2$. In Sec.~\ref{se} we present the results 
of our ultrafast spectroscopic investigations. Section \ref{st} contains a 
qualitative and quantitative account of the nonlinear magnetophononic 
phenomena we observe. In Sec.~\ref{sd} we discuss the consequences of our 
findings for the selective static and dynamical control of materials properties
both within and beyond quantum magnetism. 

\begin{figure*}[t]
\includegraphics[width=\linewidth]{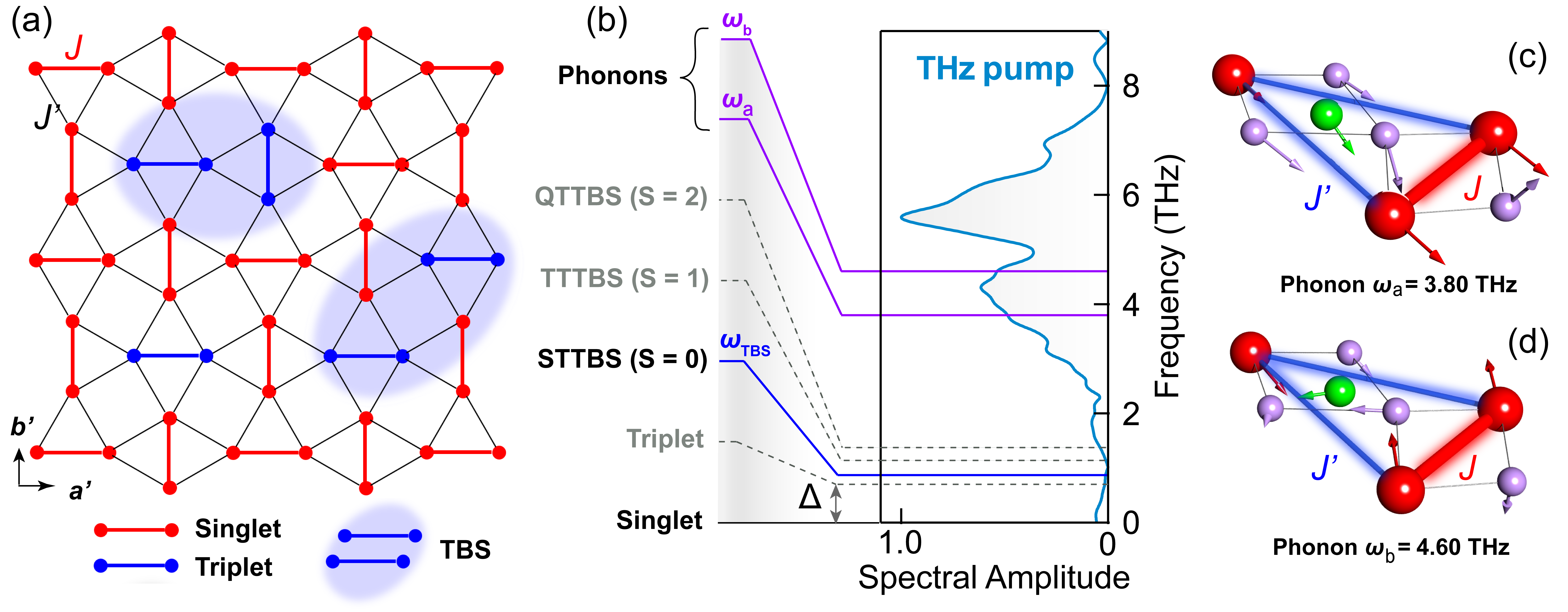}
\caption{{\bf Low-energy spin and phonon modes in SrCu$_2$(BO$_3$)$_2$.} 
(a) Schematic representation of the spin network in SrCu$_2$(BO$_3$)$_2$, 
showing how Cu$^{2+}$ ions ($S = 1/2$) form the Shastry-Sutherland geometry 
with interaction parameters $J$ on the Cu-Cu dimers and $J'$ between 
neighboring orthogonal dimers. The localized spin excitations above the 
singlet ground state (red dimers) are individual triplons (blue dimers) and 
two-triplon bound states (TBS, blue shaded regions). (b) Low-energy spectrum 
of SrCu$_2$(BO$_3$)$_2$ at ${\bf k} = 0$. In the spin sector, the singlet 
two-triplon bound state (STTBS) lies close to the one-triplon excitation, 
the triplet two-triplon bound state (TTTBS) has a smaller binding energy, 
and the quintet two-triplon bound state (QTTBS) lies close to the threshold 
for creating two isolated triplons (energies of the lowest TTTBS and QTTBS 
taken from Ref.~\cite{Nojiri03}). In the lattice sector, we show the 
frequencies of the two phonons excited most strongly in our experiment, 
$\omega_a = 3.80$ THz and $\omega_b = 4.60$ THz. The light blue line and 
gray shading represent for comparison the amplitude of the pump spectrum 
[Fig.~\ref{uqmf3}(a)]. (c-d) Lattice-displacement eigenvectors for the phonon 
modes at $\omega_a$ and $\omega_b$; because both phonons are $E$-symmetric, we 
show one of the two degenerate modes in each case. $J$ and $J'$ depend on the 
bond lengths and angles in the superexchange paths involving the Cu, O, and B 
atoms.}
\label{uqmf2}
\end{figure*}

\section{Shastry-Sutherland Model and 
S\lowercase{r}C\lowercase{u}$_2$(BO$_3$)$_2$}
\label{sscbo}

The Shastry-Sutherland model, for $S = 1/2$ spins with Heisenberg interactions 
on the two-dimensional orthogonal-dimer network shown in Fig.~\ref{uqmf2}(a) 
\cite{Shastry81}, is one of the most intriguing in quantum magnetism. The exact 
and entirely nonmagnetic ground state of singlet quantum dimers is found when 
the ratio of interdimer ($J'$) to intradimer ($J$) interactions satisfies 
$\alpha = J^\prime / J \le 0.675$, above which the QPT occurs to the 
plaquette-singlet state \cite{Corboz13}. It is quite remarkable that this 
simple model is realized so faithfully in the compound SrCu$_2$(BO$_3$)$_2$ 
[Fig.~\ref{uqmf1}(a)] \cite{Miyahara03}, with $J$ determined by Cu-O-Cu 
superexchange processes on the Cu$^{2+}$ dimer units and $J'$ by 
superexchange through the BO$_3$ units. The magnetic excitation 
spectrum of SrCu$_2$(BO$_3$)$_2$, depicted for wave vector ${\bf k} = 0$ 
in Fig.~\ref{uqmf2}(b), contains as its lowest mode the ``triplon'' 
(singlet-triplet) excitation at $\Delta = 2.9$ meV ($\equiv 0.71$ THz), 
whose dispersion is almost flat in ${\bf k}$ \cite{Gaulin04} as a consequence 
of the ideal frustration. The lattice geometry is also responsible for an 
anomalously strong binding energy for triplon pairs, with the result that 
the singlet two-triplon bound state (STTBS), the $S = 0$ branch of this 
multiplet, appears just above the one-triplon mode, at 3.6 meV ($\omega_{\rm TBS} 
 = 0.87$ THz). At higher energies, additional discrete and continuum 
excitations include the triplet ($S = 1$, TTTBS) and quintet ($S = 2$, 
QTTBS) branches of this bound state. 

These modes have been studied by a combination of neutron scattering 
\cite{McClarty17}, which targets the triplon and TTTBS, and Raman 
\cite{Lemmens00,Gozar05}, IR \cite{Room00}, and electron spin resonance 
(ESR) spectroscopies \cite{Nojiri03}, which observe the spectrum at ${\bf k}
 = 0$. To date, a detailed explanation for the phonon-assisted coupling of 
light to the spin excitations observed by Raman and IR remains elusive due 
to the inherently incoherent nature of these experiments. The interaction 
ratio in SrCu$_2$(BO$_3$)$_2$, $\alpha = 0.63$, lies close to the QPT of the 
Shastry-Sutherland model, allowing this transition to be induced under 
pressure \cite{Zayed17}. Recent attention has focused on how the magnetic 
interactions depend on the geometry of the dimer units \cite{Radtke15,
Bettler20,Badrtdinov20}, making SrCu$_2$(BO$_3$)$_2$ a fascinating and 
timely candidate for exploring ideally frustrated quantum magnetism on 
ultrafast timescales. 

\begin{figure*}[t]
\includegraphics[width=0.96\linewidth]{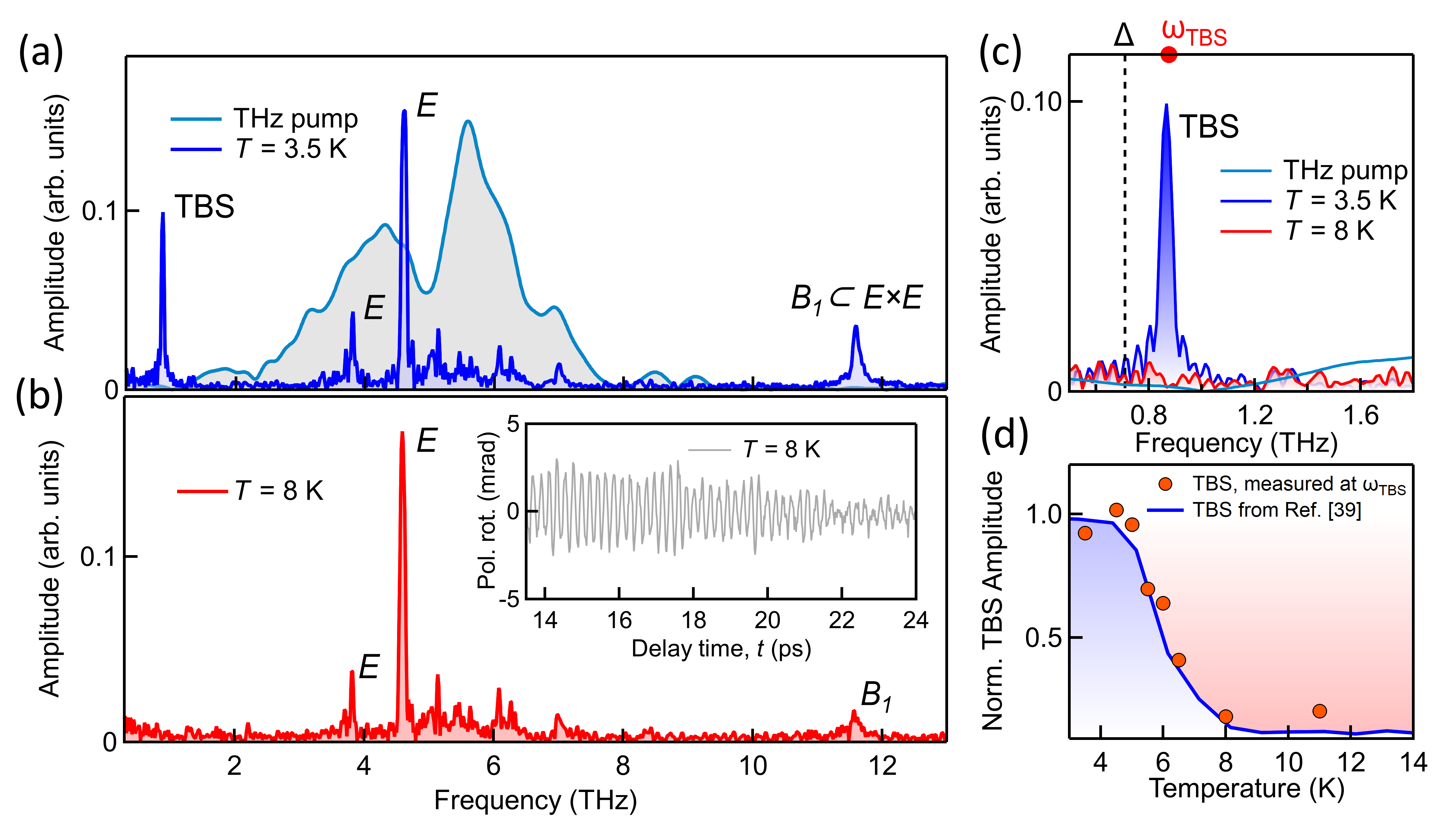}
\caption{{\bf THz field-driven lattice and spin dynamics in the frequency 
domain.} (a) Spectral amplitude (blue) of the data of Fig.~\ref{uqmf1}(c) 
computed for $3.5 \le t \le 20$ ps. The primary peaks are (i) $E$-symmetric 
phonons at 3.80 THz and 4.60 THz, (ii) a $B_1$-symmetric Raman phonon at 11.6 
THz, and (iii) the TBS excitation at $\omega_{\rm TBS} = 0.87$ THz. The light 
blue line and gray shading show the spectral amplitude of the driving electric 
field [Fig.~\ref{uqmf1}(b)]. (b) Spectral amplitude measured at 8 K, where the 
TBS is absent and the driven dynamics reveal no coherent oscillations at 
$\omega_{\rm TBS}$ (inset). (c) Comparison of low-frequency spectra at 3.5 and 
8 K. The black dashed line marks the one-triplon gap, $\Delta = 2.9$ meV 
(0.71 THz), and the red dot $\omega_{\rm TBS}$ from Raman spectroscopy 
\cite{Lemmens00,Gozar05}. (d) Temperature-dependence of the TBS, normalized 
to the peak height.}
\label{uqmf3}
\end{figure*}

\section{Ultrafast Experiment}
\label{se}

We perform THz-pump, optical-probe spectroscopy using the apparatus of 
Ref.~\cite{Vicario20}, whose technical specifications are detailed in 
App.~A. Our experiments use 
a single-crystal sample of SrCu$_2$(BO$_3$)$_2$ maintained at 3.5 K ($k_{\rm B} 
T \ll \Delta$), a temperature where the ground state is close to the pure 
singlet state. As represented in Fig.~\ref{uqmf1}(a), intense light pulses 
[Fig.~\ref{uqmf1}(b)] with spectral content between 2 and 7 THz 
[Fig.~\ref{uqmf2}(b)] drive the resonant excitation of dipole-active phonon 
modes \cite{Homes09}. To estimate the external THz electric-field strength of 
the pump, we measured the energy per pulse, beam waist, and pulse duration, 
also reported in App.~A, to obtain the value $E_{\rm THz} = 3.2$ MVcm$^{-1}$, 
which is comparable to other modern high-intensity sources \cite{Liu17,
Agranat18}. To probe the driven lattice and spin dynamics, we measured the 
ultrafast polarization rotation \cite{Nova17}, shown in Fig.~\ref{uqmf1}(c), 
caused by the associated optical birefringence and Faraday effects. These are 
imprinted on a co-propagating NIR pulse (50 fs, wavelength 800 nm) with a 
variable the delay time, $t$, and an analysis of the complete time-frequency 
response function is presented in App.~B.

As the inset of Fig.~\ref{uqmf1}(c) makes clear, the ps pump pulse creates 
dynamical oscillations that persist for many tens of ps. We take this 
observation as an opportunity to establish our use of terminology. It is clear 
that all of the physical phenomena revealed by ultrafast pulsed driving are 
transient, in the sense that the stimulus is removed after a very short time 
(here 1 ps) and the system relaxes back to equilibrium. This relaxation is 
governed by the characteristic decay times of the multiple eigenmodes of the 
system, all of which are in principle excited by the wide range of frequencies 
present in the short pulse. It is during this time that we apply the term 
``nonequilibrium,'' to refer to the situation where the system is in an 
out-of-equilibrium state, with a nonequilibrium (nonthermal) occupation of 
the states in the spectrum and the possibility of changes to the eigenenergies 
and eigenstates of this spectrum (i.e.~away from their equilibrium properties). 
In the following we will show that Fig.~\ref{uqmf1}(c) presents a particularly 
clear separation of timescales, by which the phonon modes driven by the 
electric field of the THz pulse have largely decayed after 20 ps, but the 
slow oscillation they drove, which is the leading excitation in the magnetic 
sector, lives far longer. This consequence of the very weak coupling of spins 
to their environment provides an excellent illustration of our rationale for 
studying quantum magnetism.

Returning to our experimental observations, in the frequency domain we 
find coherently excited phonons close to the peak of the pump spectrum 
[Fig.~\ref{uqmf3}(a)], primarily two of the $E$-symmetric modes measured by 
IR spectroscopy \cite{Homes09}, centered at $\omega_a = 3.80$ and $\omega_b = 
4.60$ THz. As noted above, these phonon frequencies lie far above the primary 
features in the magnetic spectrum. Nonetheless, we observe a striking response 
precisely at $\omega_{\rm TBS} = 0.87$ THz [Fig.~\ref{uqmf3}(b)], even though the 
spectral content of the pump is negligible in this frequency range. In the same 
way, the feature centered at 11.6 THz lies much higher than the spectral 
content of the pump, and the appearance of this Raman-active $B_1$ phonon mode 
in the measured response indicates that nonlinear phonon mixing is allowing 
sum-frequency excitation processes. In fact this feature constitutes one of 
the clearest examples of a sum-frequency phonon excitation yet observed, and 
thus we analyze it in detail in App.~C, but from the standpoint of 
magnetophononics it serves only as an indicator of the mechanism 
for the phenomena we investigate.

Henceforth we refer to the STTBS, whose nonmagnetic ($S = 0$) character makes 
it the only low-energy mode one may expect to excite strongly with phonons, 
simply as the TBS. To identify the excitation at 0.87 THz as the TBS, we 
undertook a number of tests. First we repeated the experiment at 8 K 
[Figs.~\ref{uqmf3}(b-c)], where thermal fluctuations cause the triplons 
to lose their character \cite{Zayed14}. While the spectral features of 
the lattice remain almost unchanged, the magnetic fingerprint of THz driving 
has disappeared. As Fig.~\ref{uqmf3}(d) makes clear, the measured amplitude 
shows rapid quenching above 5 K, exactly as observed for the TBS in 
Ref.~\cite{Lemmens00}, where this behavior was attributed to strong 
scattering from thermally excited triplets. The loss of triplon spectral 
weight at a temperature so anomalously low in comparison with the 34 K gap 
[Fig.~\ref{uqmf2}(d)] was explained only recently by the massive proliferation
of bound and scattering states of two triplons as their energy increases
\cite{Wietek19}, and the exact coincidence of the triplon \cite{Zayed14} 
and TBS decay functions indicates the origin of the behavior we observe 
in Fig.~\ref{uqmf3}(d) \cite{Lemmens00}. The reduced lifetime of the 
$B_1$ phonon in Fig.~\ref{uqmf3}(b) also provides evidence of damping 
processes due to spin-lattice coupling.

\begin{figure}[t]
\includegraphics[width=\columnwidth]{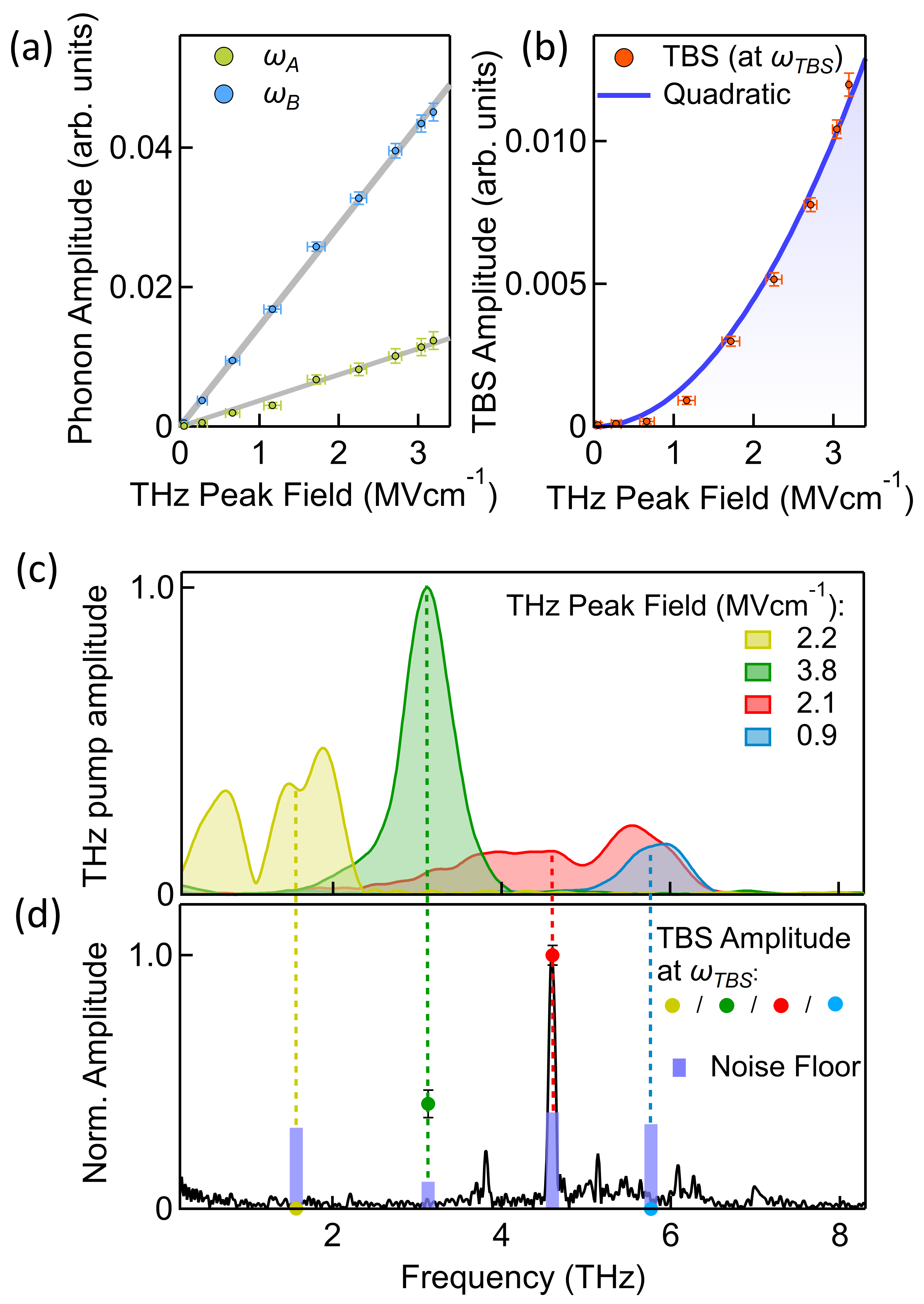}
\caption{{\bf THz pump strength and frequency-dependence.} (a) Linear 
dependence of phonon amplitudes and (b) quadratic dependence of the TBS 
amplitude on the electric-field strength. (c) Four different pump spectra 
obtained by filtering of the THz beam and labelled by their peak field values 
in the time domain; dashed lines mark the centroid of each spectral envelope. 
(d) TBS mode amplitude (colored points), normalized to the square of the peak 
field, for the four different values of the THz pump-frequency centroid in 
panel (c). Bars mark the noise floor of the extracted signal in each case.}
\label{uqmf4}
\end{figure}

To understand the mechanism driving a purely magnetic excitation, the fact 
that the spectral content of the pump contains negligible intensity below 
2 THz excludes a direct coupling of light to the TBS, which in any case is 
not IR-active \cite{Room00}. To probe the indirect origin of the observed 
spin dynamics, we measure the mode amplitudes as functions of the THz 
electric-field strength. These amplitudes and their uncertainties are 
extracted from Gaussian fitting as described in App.~D.
The IR phonons display the linear dependence expected for resonant excitation 
[Fig.~\ref{uqmf4}(a)]. By contrast, the TBS amplitude varies quadratically 
with the field strength [Fig.~\ref{uqmf4}(b)], and thus with lattice 
displacement, indicating a nonlinear coupling mechanism essentially 
different from electric or magnetic dipolar interactions \cite{Kampfrath11,
Schlauderer19}. Confirmation that this dependence is quadratic can also be 
obtained from the invariance of the output signal on inverting the polarity 
of the pump electric field, as we show in App.~E.

For a direct demonstration of which IR phonons provide the driving, 
we vary the spectral content of the pump by the insertion of different 
high- and low-pass filters, as described in App.~A. The resulting pump 
spectra are shown in Fig.~\ref{uqmf4}(c) and again the corresponding TBS 
amplitude, shown in Fig.~\ref{uqmf4}(d), is large only when the pump 
spectrum covers $\omega_a$ and $\omega_b$. There is no sign of coherent 
magnetic dynamics in off-resonant conditions, as we show explicitly for 
driving frequencies predominantly below 2 and above 5.5 THz. 

This proves that the opto-magnetic coupling is mediated by the resonant 
lattice excitations. However, one may still suspect that the phonon driving 
is not direct, but occurs rather as a consequence of nonlinear effects 
involving non-resonant electronic excitations. THz-induced free-carrier 
generation could result in the excitation of coherent phonons by displacive 
processes \cite{Zeiger92}, with modes $\omega_a$ and $\omega_b$ being favored 
by their spectral weight, and indeed the displacive mechanism has been the 
subject of some recent ultrafast studies \cite{Jnawali21,Giorgianni22}. To 
eliminate the possibility of THz-induced free-carrier generation in 
SrCu$_2$(BO$_3$)$_2$, we measured the transmission modulation of the NIR 
probe pulse whose polarization rotation is shown in Fig.~\ref{uqmf1}(c). 
The results of this study, which we show in App.~F, 
demonstrate conclusively that no free carriers are produced by our THz pump 
pulse, and hence that any mechanisms generating these, which include 
THz-induced electronic breakdown \cite{Yamakawa17} or impact ionization 
\cite{Tarekegne17,Hubmann20}, are excluded in SrCu$_2$(BO$_3$)$_2$.

Focusing now on the direct driving of coherent phonons, for a quantitative 
analysis of the effect of the THz pulse on the atomic motions within the 
sample, we equate the polarization induced by the electric field with the 
modulation of the dipole moment, which peaks strongly when the pulse 
frequency is resonant with a phonon mode, denoted by $m$. In this 
situation $P_m = n_d \delta_m \mu_m$, where $\mu_m$ is the net charge 
displacement due to mode $m$, $n_d$ is a dipole density, and $\delta_m$ is 
the maximum displacement coordinate of the phonon. In this way we deduce 
(App.~A) maximum displacements up to $\delta_b = 0.17$ 
\AA~for the 4.60 THz mode [shown in Fig.~\ref{uqmf2}(d)], which is comparable 
to the value estimated in SrTiO$_3$ \cite{Kozina19}. We stress that $\delta_m$ 
represents the maximum displacement of the most displaced O ion in the 
SrCu$_2$(BO$_3$)$_2$ structure due to phonon mode $m$, and that the 
corresponding displacements of the Cu ions are generally smaller (by a 
factor of 3-4 for the 4.60 THz mode), whence the system does not approach 
the Lindemann melting criterion. Because the temporal duration of the pump 
pulse is approximately 0.5 ps [Fig.~\ref{uqmf1}(b)], the amount of energy 
transferred presents little danger that the THz driving can increase the 
sample temperature significantly. 

\begin{figure*}[t]
\includegraphics[width=\linewidth]{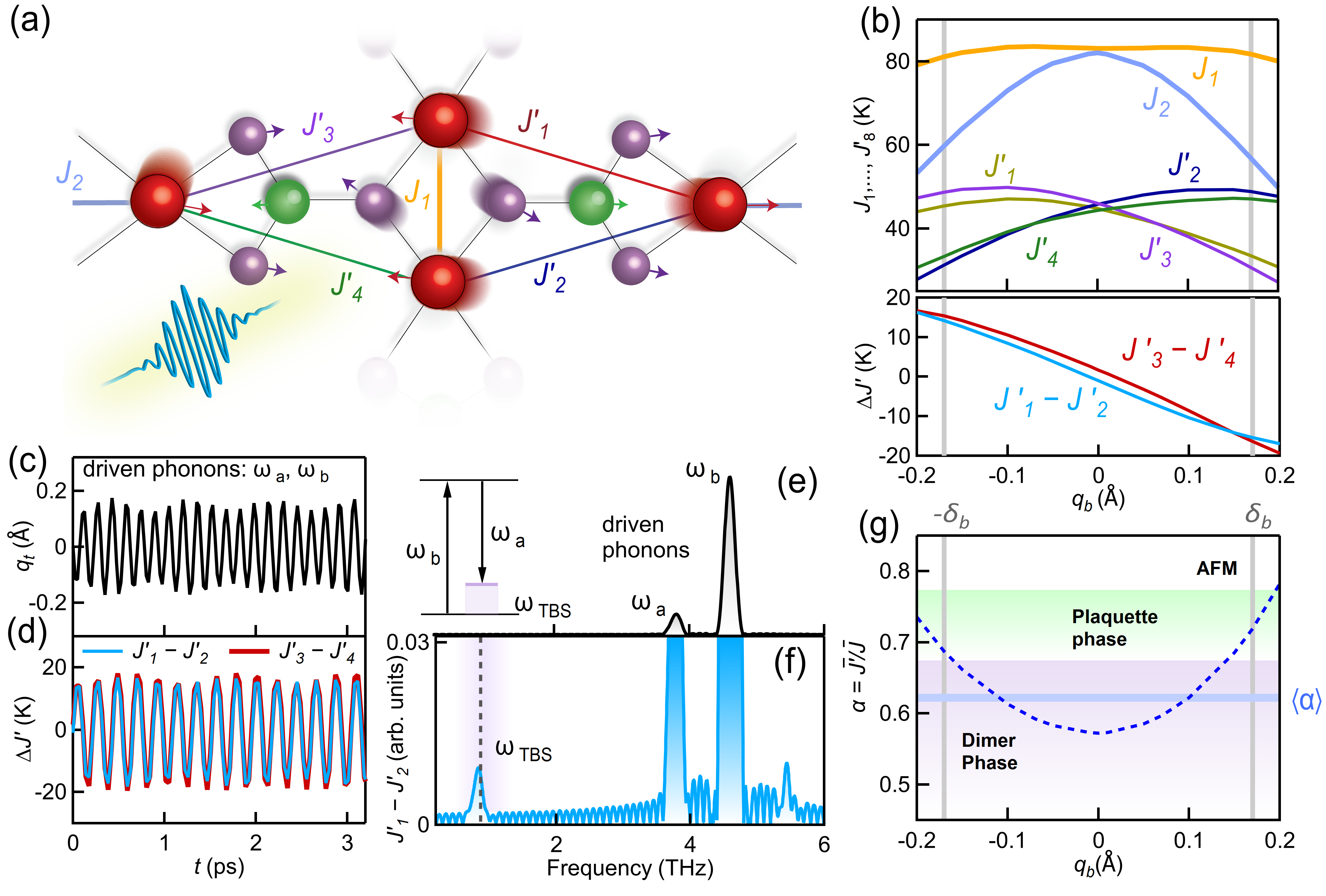}
\caption{{\bf Dynamic control of magnetic interactions by light-driven 
phonons.} (a) Schematic representation of atomic motions in the $ab$-plane 
associated with the $\omega_b = 4.60$ THz phonon. (b) Interaction parameters 
$J$, $J_1'$, $J_2'$, $J_3'$, $J_4'$, $\Delta J_{12}'$, and $\Delta J_{34}'$ 
calculated for the symmetry-broken lattice structure as functions of the 
corresponding phonon displacement, $q_b$. (c) Temporal modulation of $q_a
 + q_b$ calculated using the experimental THz driving field. (d) Corresponding 
dynamic modulation of $\Delta J_{12}'$ and $\Delta J_{34}'$, which are controlled 
primarily by the $\omega_b$ phonon. (e) Spectrum of driven IR phonons, which 
shows weight only at the frequencies $\omega_a$ and $\omega_b$. Inset: 
energy-level diagram of the TBS excitation process, which is a resonance 
between the phonon frequency difference, $\omega_b - \omega_a$, and 
$\omega_{\rm TBS}$. (f) Fourier transform of the DFT time series for $\Delta 
J_{12}' (t)$, which shows spectral weight at $\omega_{\rm TBS}$. (g) Interaction 
ratio, $\alpha = \bar{J}'/\bar{J}$, where $\bar{J} = {\textstyle \frac12} (J_1
 + J_2)$ and $\bar{J}' = {\textstyle \frac{1}{8}} (J_1' + ... + J_8')$, shown 
as a function of $q_b$. The vertical lines indicate the maximum value of $q_b$ 
estimated from our peak electric field and the horizontal line shows the 
time average, denoted $\langle \alpha \rangle$, of the interaction ratio 
at this driving field.}
\label{uqmf5}
\end{figure*}

\section{Theory: Frustration-Breaking}
\label{st}

\subsection{Analysis}

We now establish the physical framework for the nonlinear magnetophononic 
phenomenon we have created. Qualitatively, the lattice displacements due 
to any phonon excitation alter the instantaneous magnetic interactions 
[Fig.~\ref{uqmf5}(a)]. To analyze this situation we express the Hamiltonian 
of the driven system as
\begin{equation} 
H = H_0 + H_H + V_{\rm Ph} + H_{\rm Ph-Em} + H_D,
\label{edh} 
\end{equation} 
where $H_0$ is spin-independent and 
\begin{equation}
H_H = J \sum_{\langle ij \rangle} {\vec S}_i \cdot {\vec S}_j
 + J' \sum_{\langle\langle ij \rangle\rangle} {\vec S}_i \cdot {\vec S}_j ,
\label{ehh}
\end{equation}
is the Shastry-Sutherland model at equilibrium: ${\vec S}_i$ is a spin-1/2 
operator located on the Cu$^{2+}$ ion at site $i$, $J$ and $J'$ are the 
magnetic interactions depicted in Fig.~\ref{uqmf2}(a), and $\langle ij 
\rangle$ and $\langle\langle ij \rangle\rangle$ denote respectively pairs 
of sites on intra- and interdimer bonds. $V_{\rm Ph}$ is the phonon potential 
and $H_{\rm Ph-Em} = \sum_m \mu_m q_m E_{\rm THz}$, the dipole coupling between the 
lattice and the THz light, is the term driving the excitation of IR-active 
phonon modes. The action of the coherent lattice deformation due to all of 
these phonon modes, $q_t = \sum_m q_m$, in modulating the magnetic interaction 
parameters causes additional coupled spin-phonon terms to enter the 
Hamiltonian, which we collect in $H_D$. To separate the phonon modulation 
terms in $H_D$, it is convenient to perform a Taylor expansion of the magnetic 
interactions, $J$ and $J'$, in powers of the simultaneously driven coherent 
IR phonon coordinates, which we denote as $q_m$, $q_n$, \dots Denoting an 
arbitrary interaction term as ${\tilde J}$, we obtain
\begin{eqnarray} 
{\tilde J} (q_m, q_n, \dots) & = & {\tilde J} (0) + \left. \frac{\partial 
{\tilde J}}{\partial q_m} \right|_{q_m = 0} q_m \label{ete} \\ & & \;\;\;\; 
\;\;\;\;  + \left. \frac{\partial^2 {\tilde J}}{\partial q_m \partial q_n} 
\right|_{q_m,q_n = 0} q_m q_n + \dots \nonumber
\end{eqnarray} 
where the first term is part of $H_H$ and the higher terms make clear the 
direct dependence of the additional contributions on the oscillating phonon 
coordinates.

Both the key features of our experiment become clear immediately in a minimal 
model with only two harmonic IR phonons, i.e.~$q_t = q_a + q_b$, where $q_a$ 
and $q_b$ are vectors of normal-mode coordinates with cosinusoidal time 
structures at respective frequencies $\omega_a$ and $\omega_b$. The $q_t^2$ 
terms appearing in the second line of Eq.~(\ref{ete}) lead to spectral 
components at frequencies $2\omega_a$, $\omega_a + \omega_b$, $2\omega_b$, 0, 
and $\omega_b - \omega_a$. These quadratic terms provide the leading nonlinear 
mechanism that allows a very wide range of spin excitation energies to be 
addressed using the sum and difference frequencies of driven IR phonons, 
whose frequency range may be more restricted. 

The second key feature is that the unconventional physics of the 
Shastry-Sutherland model, and by extension of SrCu$_2$(BO$_3$)$_2$ at 
equilibrium, relies on the ideal frustration of the spin correlations between 
dimers. The term ${\vec S}_i \! \cdot \! {\vec S}_j$ acting on a single dimer 
is an eigenoperator and when acting between dimers it is exactly cancelled 
by a second ${\vec S}_i \! \cdot \! {\vec S}_j$ term with equal size ($J'$) 
and effectively opposite sign [Fig.~\ref{uqmf2}(a)]. However, the excited 
phonons driving the lattice out of equilibrium cause frustration-breaking in 
the magnetic sector, in the form of a finite interdimer coupling, $\Delta 
J_{12}' = J_1' (q_m, q_n, \dots) - J_2' (q_m, q_n, \dots)$ (and similarly for 
$\Delta J_{34}'$, $\Delta J_{56}'$, and $\Delta J_{78}'$), that connects 
nearest-neighbor dimers [Fig.~\ref{uqmf5}(a)]. This has the immediate 
effect of allowing two qualitatively new types of physical process that 
are forbidden at equilibrium. 

To make these most transparent, we reexpress the out-of-equilibrium spin 
Hamiltonian in terms of triplet creation and annihilation operators, which 
represent the excitations of the dimer ($J$) units in Fig.~\ref{uqmf2}(a) above 
their singlet ground state. A full discussion of the bond-operator description 
may be found in Ref.~\cite{Normand11}. Treating the singlets as a scalar term 
gives conventional leading-order triplet processes of the form
\begin{eqnarray} 
H_D & = & \sum_i [\Delta J_{12}' (t_{1,i}^\dag t_{2,i+x} + t_{1,i}^\dag t_{2,i+x}^\dag)
\nonumber \\ & & \;\;\;\;\;\; + \Delta J_{34}' (t_{2,i}^\dag t_{1,i+x} + t_{2,i}^\dag 
t_{1,i+x}^\dag) \label{ettc}\\ & & \;\;\;\;\;\; + \Delta J_{56}' (t_{1,i}^\dag 
t_{2,i+y} + t_{1,i}^\dag t_{2,i+y}^\dag) \nonumber \\ & & \;\;\;\;\;\; + \Delta J_{78}'
 (t_{2,i}^\dag t_{1,i+y} + t_{2,i}^\dag t_{1,i+y}^\dag  )] + {\rm H.c.}, \nonumber
\end{eqnarray}
where the full set of eight inequivalent $J'$ bonds is shown in 
Fig.~\ref{uqmfs7} of App.~G. This may be contrasted with the explicit treatment 
for SrCu$_2$(BO$_3$)$_2$ at equilibrium in Ref.~\cite{McClarty17}, where all of 
these bilinear triplon terms vanish. The first term in each bracket describes 
the propagation of existing triplon excitations between dimers, relieving 
their strict localization, although the high-frequency oscillation of $\Delta 
J'$ does not allow any quasi-static changes of the flat triplon bands. 

The second term in Eq.~(\ref{ettc}) describes two-triplon creation on 
neighboring dimer pairs, i.e.~the direct excitation of the TBS from the 
singlet quantum ground state ($|s \rangle$). While the triplon pairs are 
created initially on adjacent dimers, on the timescale of the spin system 
they will optimize their relative configuration to form the most strongly 
bound state \cite{McClarty17}, which is depicted very schematically on the 
right side of Fig.~\ref{uqmf2}(a). Because lattice excitations cannot change 
the spin quantum numbers ($\Delta S_{\rm tot} = 0$) in a spin-isotropic 
Hamiltonian, only transitions to the STTBS ($S = 0$) are allowed. The 
coefficients $\Delta J_{12}'$, \dots $\Delta J_{78}'$ in Eq.~(\ref{ettc}) 
are a direct expression of the frustration-breaking, making fully explicit 
the origin of this phonon-driven triplon pair creation.

\subsection{Lattice dynamics and density functional theory}

We perform two types of quantitative calculation within this framework. 
To model the nonlinear effects of the driven phonons, we follow 
Refs.~\cite{Foerst11} and \cite{Melnikov18} by considering $H_D (q_m, q_n)$ 
as a dynamic perturbation through the coupling term $H_{\rm Em-Ph}$ in 
Eq.~(\ref{edh}). In our experiments (Sec.~\ref{se}), two different IR-active 
phonon modes show the strongest driving. Labelling these by $a$ and $b$, 
their equations of motion are 
\begin{eqnarray}
{\ddot q}_a + \gamma_a {\dot q}_a + \omega_a^2 q_a & = & - B_a E_{\rm THz} (t), 
\nonumber \\
{\ddot q}_b + \gamma_b {\dot q}_b + \omega_b^2 q_b & = & - B_b E_{\rm THz} (t),
\label{edpd}
\end{eqnarray}
where $\omega_a = 3.80$ THz and $\omega_b = 4.60$ THz are the experimental 
phonon frequencies, $\gamma_a = 0.02$ THz and $\gamma_b = 0.03$ THz are 
their respective damping rates (App.~B), and $E_{\rm THz}$ is the external 
THz driving field taken from experiment [Fig.~\ref{uqmf1}(b)]. $B_a$ and 
$B_b$ are the dipolar coupling constants, which depend on the effective 
charges ($Z_{\rm eff}^i$), the transmission coefficients ($\beta_m$), and the 
reduced masses of the phonon modes, which we deduce from the maximum 
displacements $\delta_a = 0.04$ \AA~and $\delta_b = 0.17$ \AA~calculated 
following App.~A. From Eq.~(\ref{edpd}) we compute as a function of time 
the net atomic displacement due to the two leading phonon modes, $q_t = 
q_a + q_b$, obtaining the result shown in Fig.~\ref{uqmf5}(c). 

The second type of calculation is to compute the magnetic interaction 
parameters by DFT. These calculations were performed using the Quantum 
Espresso package \cite{Giannozzi17}, an open-source tool for electronic 
structure calculations based on DFT and the pseudopotential plane-wave 
technique. Exchange and correlation effects were modelled using the PBE 
functional \cite{Perdew96}, augmented by a Hubbard $U$ term to include the 
strongly correlated nature of the Cu 3$d$ electrons. The calculation of 
magnetic interactions is a self-consistent process in which the lattice 
structure is relaxed fully in a selected collinear spin configuration for each 
fixed value of the effective $U$ parameter and then the total energies of the 
different magnetic configurations are compared \cite{Radtke08} by mapping them 
onto the terms $H_0 + H_H$ in Eq.~(\ref{ete}) to determine equilibrium values 
for $J$ and $J'$ [Eq.~(\ref{ehh})]. The full details of this process and of 
its intrinsic accuracy are presented in App.~G.

In the initial step of our calculations we used this procedure to refine $U$, 
deducing that $U = 11.4$ eV yields the magnetic interactions $J = 7.24$ meV 
(84.0 K) and $J' = 4.28$ meV (49.7 K), in good agreement with experimental 
findings \cite{McClarty17} and giving a coupling ratio $\alpha = 0.592$ 
for the equilibrium lattice structure. We then extended these methods to 
estimate the phonon-induced modulation of the magnetic interaction parameters 
by computing the temporal evolution of the frustration-breaking terms, $\Delta 
J'(t)$. For this we evaluated the magnetic interactions in a dense sequence of 
different ``frozen phonon'' configurations of the lattice. Each atom in the 
SrCu$_2$(BO$_3$)$_2$ structure was displaced by $q_m {\hat u}_{im}$, where $q_m$ 
is the instantaneous displacement amplitude of excited phonon mode $m$, ${\hat 
u}_{im}$ denotes the set of normal-mode vectors taken from Ref.~\cite{Homes09}, 
and we restricted our calculations to $m = a,b$ with $\omega_a = 3.80$ and 
$\omega_b = 4.60$ THz. The displacements of the atoms from equilibrium reduce 
the lattice symmetry, as represented in Fig.~\ref{uqmf5}(a), and hence require 
more complex calculations of more interaction parameters in a larger unit cell, 
as detailed in App.~G.

In Fig.~\ref{uqmf5}(b) we show the four different interdimer interaction 
parameters $(J_1', J_2', J_3', J_4')$ neighboring each vertical spin dimer as 
functions of the largest scalar phonon displacement amplitude, $q_b$, whose 
maximum value is $\delta_b$. We stress that the parameters $(J_5', J_6', J_7', 
J_8')$, which we do not show, are different from $(J_1', J_2', J_3', J_4')$ 
for these specific members of the $E$-symmetric phonon doublets, but are 
interchanged with them for the other member of the doublet. Figure 
\ref{uqmf5}(b) shows that, while the phonon-induced variations in $J_1$ and 
$J_2$ are almost quadratic, the interactions $J'_{1,...,4}$ all have significant 
linear components. These result in large values of the frustration-breaking 
difference interactions, $\Delta J'$, well in excess of 10 K (i.e.~reaching 
20-30\% of $J'$ by this estimate). 

To mimic our experiment, we superposed the phonon normal modes to obtain 
$q_t = q_a + q_b$ [Fig.~\ref{uqmf5}(c)] at 285 time points in steps of 0.035 
ps, in order to span 10 ps. The corresponding frustration-breaking 
interaction parameters, $\Delta J_{12}'$ and $\Delta J_{34}'$ 
[Fig.~\ref{uqmf5}(d)], show a strong and in-phase oscillation controlled 
largely by $q_b$. In the frequency domain, this density of time steps 
reproduces well-resolved peaks at both the low frequency of the TBS peak 
and the intermediate frequency of the driven phonons. Even at the harmonic 
level we model, the IR phonons excited by the THz pump [Fig.~\ref{uqmf5}(e)] 
create a nonlinear modulation of the interaction parameters [through 
Eq.~\eqref{ete}] with a rich spectrum [Fig.~\ref{uqmf5}(f)]. This spectrum 
includes significant weight at the frequency $\omega_b - \omega_a$, whose 
resonance with the TBS [inset, Fig.~\ref{uqmf5}(e)] ensures large values of 
the matrix element for two-triplon creation, $\langle {\rm TBS} | \Delta J' 
(\omega) t_{1,i}^\dag t_{2,j}^\dag |s \rangle$, and hence the strong nonequilibrium 
population of this purely magnetic excitation that we report in 
Figs.~\ref{uqmf3}(a) and \ref{uqmf3}(c).

\section{Discussion}
\label{sd}

Ideal frustration of magnetic interactions is the core attribute that leads 
quantum spin systems to form entirely unconventional and exotic phases, 
although some of the most fundamentally novel properties (such as fractional 
excitations, topological order, and entanglement) often remain largely hidden 
from conventional experimental probes. In this context, a static breaking of 
ideal frustration is often regarded as trivial, merely removing the properties 
that make the system special; in SrCu$_2$(BO$_3$)$_2$, a static $\Delta J'$ 
would restore triplon propagation, destabilize the bound states, and favor 
the nearby phase of long-ranged antiferromagnetic order \cite{Corboz13}. 
However the dynamical frustration-breaking we effect means that the static 
and equilibrium state of ideal frustration is retained, excluding most of 
these more trivial hallmarks and thereby offering an alternative route to 
the selective and controlled investigation of certain hidden properties.

While much has been made of ``controlling quantum systems'' on ultrafast 
timescales, we stress that ultrafast processes in magnetic materials have to 
date been restricted in large part to destroying, modulating the magnitude, or 
switching the direction of an ordered moment \cite{Kampfrath11,Foerst15,Disa20}.
Here we have solved two further fundamental problems on the route to ultrafast 
dynamical control. The first is the coupling of light to a quantum magnet with 
no magnetic order. As a lattice-based route, magnetophononics achieves such 
coupling both universally, meaning for all forms of interaction in condensed 
matter -- however ``hidden'' the resulting unconventional order may be, or how 
``forbidden'' a process may be, in experiments performed at equilibrium -- and 
in an intrinsically resonant manner. The second is the frequency-matching 
problem by which the light-driven pump (the phonons) can be tuned to the target 
(magnetic) excitations, and for this purpose our second-order nonlinear variant 
of magnetophononics allows a universal extension of lattice control to very low 
(phonon difference) and high (phonon sum) frequencies. 

At the conceptual level, by periodic driving of a quantum system one may expect 
to achieve state control, population control, or spectral control, all of 
which are covered in the term ``Floquet engineering'' \cite{Oka19}. In the 
true Floquet regime, where the frequency of the light far exceeds the system 
frequencies, driving causes no significant modification of the basis states 
(very little energy is transferred), but induces phase-coherent superpositions 
of these states that can have dramatically different properties. When the 
light frequency resonates with a system transition, the high available laser 
intensities allow one to achieve strongly nonequilibrium state populations, 
raising the prospect of driving a nonequilibrium Bose condensation in 
suitable systems. While these processes may alter the spectrum of the system 
as a secondary effect, a direct modulation of the system parameters can be 
achieved in one of two ways. A simple situation would be the opposite limit 
to SrCu$_2$(BO$_3$)$_2$, where slow phonons in a material with high magnetic 
energies give linear control of the parameters. However, magnetophononic 
driving at the resonant frequencies of the spin system offers the clearest 
potential for unconventional dynamical phenomena, including qualitatively 
different types of hybrid spin-phonon state, engineering of the spin and 
phonon energies, and strong mutual feedback effects on state populations 
\cite{rykun21}. 

In SrCu$_2$(BO$_3$)$_2$ we have achieved control in the form of creating a 
significant nonequilibrium population of a target excited state. However, 
this population was not sufficient, at our available electric fields, to 
cause a detectable modification of the TBS energy. In the limit exemplified 
by SrCu$_2$(BO$_3$)$_2$, where the phonons are much faster than the spin 
excitations and thus resonant driving is caused by a rather weak 
second-harmonic component, we have not yet been able to alter the TBS or 
triplon energies directly. However, considering the many different phonon 
normal modes and frequencies available in SrCu$_2$(BO$_3$)$_2$, it is clear 
that such dynamical driving presents a wide variety of highly specific control 
channels for different interactions in the atomic structure, in stark contrast 
to the universal alterations effected by conventional static control channels, 
such as an applied pressure or magnetic field. Here we recall also that the 
pulsed nature of the experimental driving, while essential for heat control 
in the sample, implies that all of the phenomena observed will be long-lived 
transients constituting an approximation to the true nonequilibrium states 
of the driven system. 

As a more general route to nonresonant interaction control, Fig.~\ref{uqmf5}(g) 
indicates for the example of a driven $\omega_b$ phonon how the method of 
driving IR-active phonons in the harmonic regime can cause a strong increase 
in the instantaneous ratio of the spatially averaged interactions, $\alpha = 
\bar{J}'/\bar{J}$. While recent experimental \cite{Bettler20} and theoretical 
\cite{Badrtdinov20} studies have highlighted the role of the ``pantograph'' 
phonon \cite{Radtke15} in modulating $\alpha$, this $A_1$-symmetric mode, 
found at 6.1 THz in Raman measurements, cannot be excited coherently in the 
same way as IR phonons. Because the phonons are so much faster than the 
magnetic interactions, the important quantity is the time average of $\alpha 
(q_b)$ in Fig.~\ref{uqmf5}(g), whose rise with $q_b$ indicates the prospect, 
at still higher electric fields, of a driven, dynamical approach to the 
static QPT into the plaquette phase of SrCu$_2$(BO$_3$)$_2$, i.e.~with no 
frustration-breaking effects on long timescales. Although it represents 
the effect of only a single phonon, Fig.~\ref{uqmf5}(g) suggests that the 
dynamical approach to controlling $\alpha$ is clearly comparable in range 
with hydrostatic pressure techniques \cite{Haravifard16,Zayed17}; indeed our 
current studies, which were not optimized for this purpose, indicate a high 
potential for achieving significantly stronger static effects by further 
increasing the THz field strength and by tailoring of the excited phonon modes. 

Beyond the primary superexchange interactions ($J$ and $J'$), further terms 
in the modulated spin Hamiltonian can also produce spin excitations that are 
normally weak or forbidden. SrCu$_2$(BO$_3$)$_2$ has Dzyaloshinskii-Moriya (DM) 
interactions \cite{Nojiri03}, which are small (3\% of $J$), but are important 
in applied magnetic fields, including to create topological states 
\cite{McClarty17}. The driving of symmetry-breaking IR phonons also creates 
dynamical antisymmetric spin interactions directly analogous to the modulation 
of $J$ and $J'$. In SrCu$_2$(BO$_3$)$_2$, our experiments demonstrate no 
discernible role for driven DM interactions, because the one-triplon excitation 
process ($\Delta S_{\rm tot} = 1$) at 0.71 THz should be excited by the same 
difference-frequency envelope as the TBS, but clearly no feature is visible 
above the detection threshold at this frequency in Fig.~\ref{uqmf3}(c). More 
generally, however, one may use selective nonlinear IR phonon driving to 
manipulate both the symmetric and antisymmetric magnetic interactions in 
systems such as skyrmion lattices and chiral spin liquids, where both 
couplings play an essential role.

In summary, we have demonstrated coherent light-driven spin dynamics in a 
purely quantum magnetic system. By the resonant excitation of phonons and 
their nonlinear mixing to span a very wide (sum and difference) frequency 
range, our experimental protocol meets the intrinsic challenge of spin-phonon 
frequency-matching. We have applied it to SrCu$_2$(BO$_3$)$_2$ and achieved 
the selective excitation of the singlet branch of the two-triplon bound state 
without exciting individual triplons. We have shown theoretically how this 
process occurs, once the driven phonons relieve the ideal magnetic frustration, 
and have performed DFT calculations to estimate the magnitude of the 
interaction modulation. Our results open an additional time dimension 
for exploring quantum magnetic phenomena that to date have been probed 
only by quasi-static stimuli, and because it uses the lattice as its medium 
our method is applicable without restriction to all the exotic spin states 
available in quantum magnetic materials. In view of ongoing technical progress 
at all the frontiers of narrow-band spectra, high intensities, and ultrashort 
pulses, one may anticipate order-of-magnitude improvements in both driving 
and detection that will place dynamically driven phenomena within reach in 
multiple classes of quantum material.

\begin{acknowledgments} 
We are grateful to C. Homes for sharing the calculated phonon eigenvectors. 
We thank T. Cea, M. F\"orst, S. Furuya, A. Kimel, R. Mankowsky, F. Mila, A. 
Razpopov, G. S. Uhrig, and R. Valent\'{\i} for valuable discussions. This 
research was supported by the European Research Council (ERC) within the 
EU Horizon 2020 research and innovation programme under Grant No.~681654 
(HyperQC), by the MARVEL National Centre of Competence in Research of the 
Swiss National Science Foundation and by the DFG (German Research Foundation) 
through Grant No.~UH90/13-1.
\end{acknowledgments} 

\bibliography{uqm}

\onecolumngrid

\vspace{2cm}

\begin{figure*}[tbh]
\includegraphics[width=0.9\linewidth]{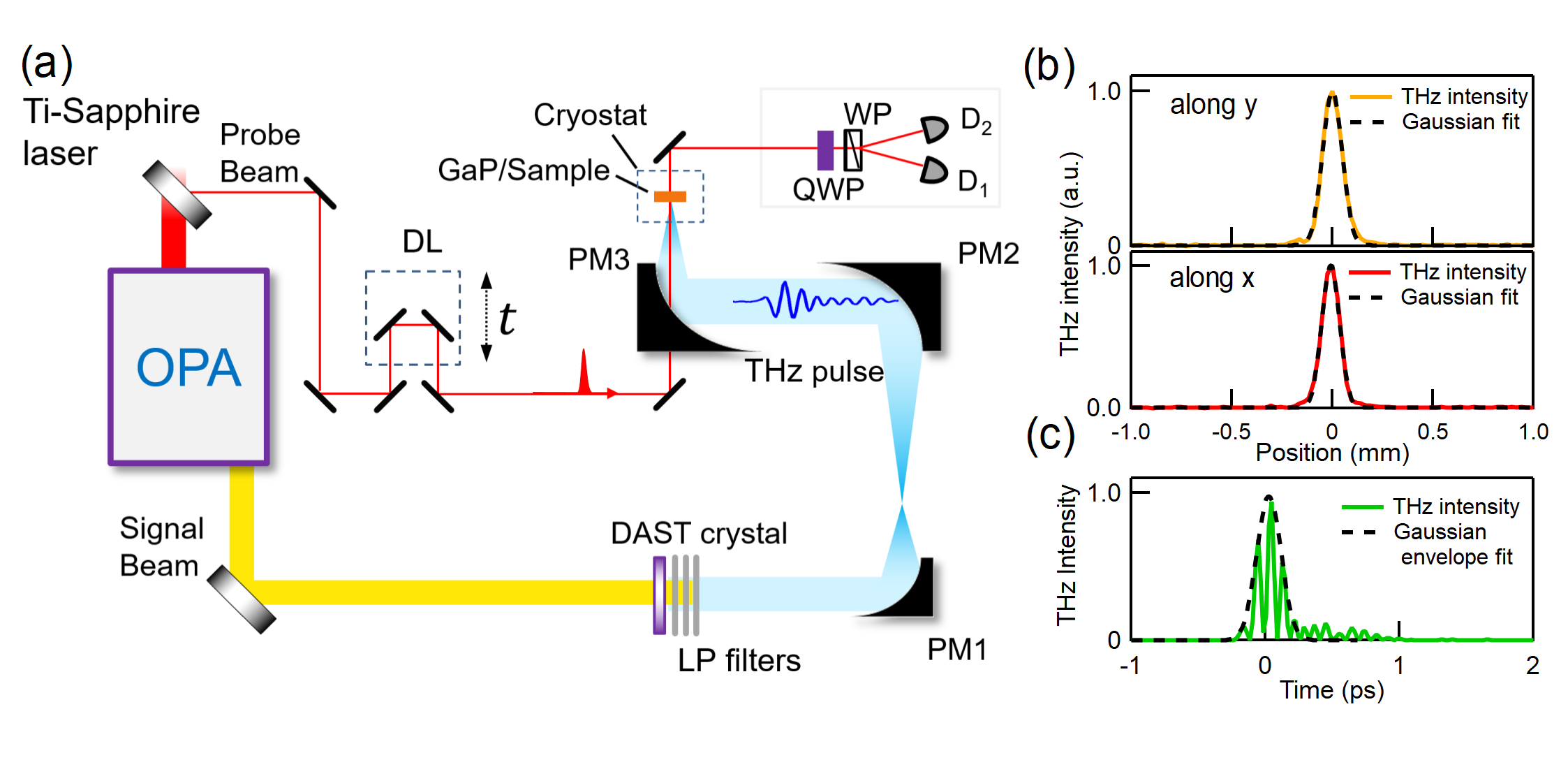}
\caption{{\bf Experimental THz-pump, optical-probe set-up and measured THz 
pump parameters.} (a) Schematic representation showing the Optical Parametric 
Amplifier (OPA), parabolic mirrors (PM1-PM3), delay line (DL), quarter-wave 
plate (QWP), Wollaston prism (WP), and detectors (D1-D2). (b) THz beam profile 
at the sample position measured by the THz camera. Beam waists obtained by 
Gaussian fitting are respectively $w_x = 88$ $\mu$m and $w_y = 96$ $\mu$m 
(average waist $w = 92$ $\mu$m). (c) Temporal THz intensity waveform obtained 
as the square of the electric field measured by electro-optic sampling, using 
a Gaussian-envelope fit to determine the pulse duration.}
\label{uqmfs1}
\end{figure*}

\twocolumngrid

\begin{appendix}

\section{Experiment}
\label{ses}

Single crystals of SrCu$_2$(BO$_3$)$_2$ were grown using an optical 
floating-zone furnace (FZ-T-10000-H-IV-VP-PC, Crystal System Corp., Japan) 
with four 300 W halogen lamps as the heat source. The growth rate was 0.25 
mm/h, with both feeding and seeding rods being rotated at approximately 15 
rpm in opposite directions to ensure the homogeneity of the liquid; an argon 
atmosphere with 20\% oxygen was maintained at 5 bar during growth. The high 
structural quality and orientation of the resulting single crystal were 
confirmed by x-ray diffraction and the high magnetic quality (absence of 
impurities) by susceptibility measurements. The crystal was cut using a 
diamond wire saw and cleaved along the $a'b'$-plane to give a sample of 
dimensions 1$\times$3$\times$0.24 mm$^3$ that was used for the experiments.

\begin{figure*}[t]
\includegraphics[width=0.9\linewidth]{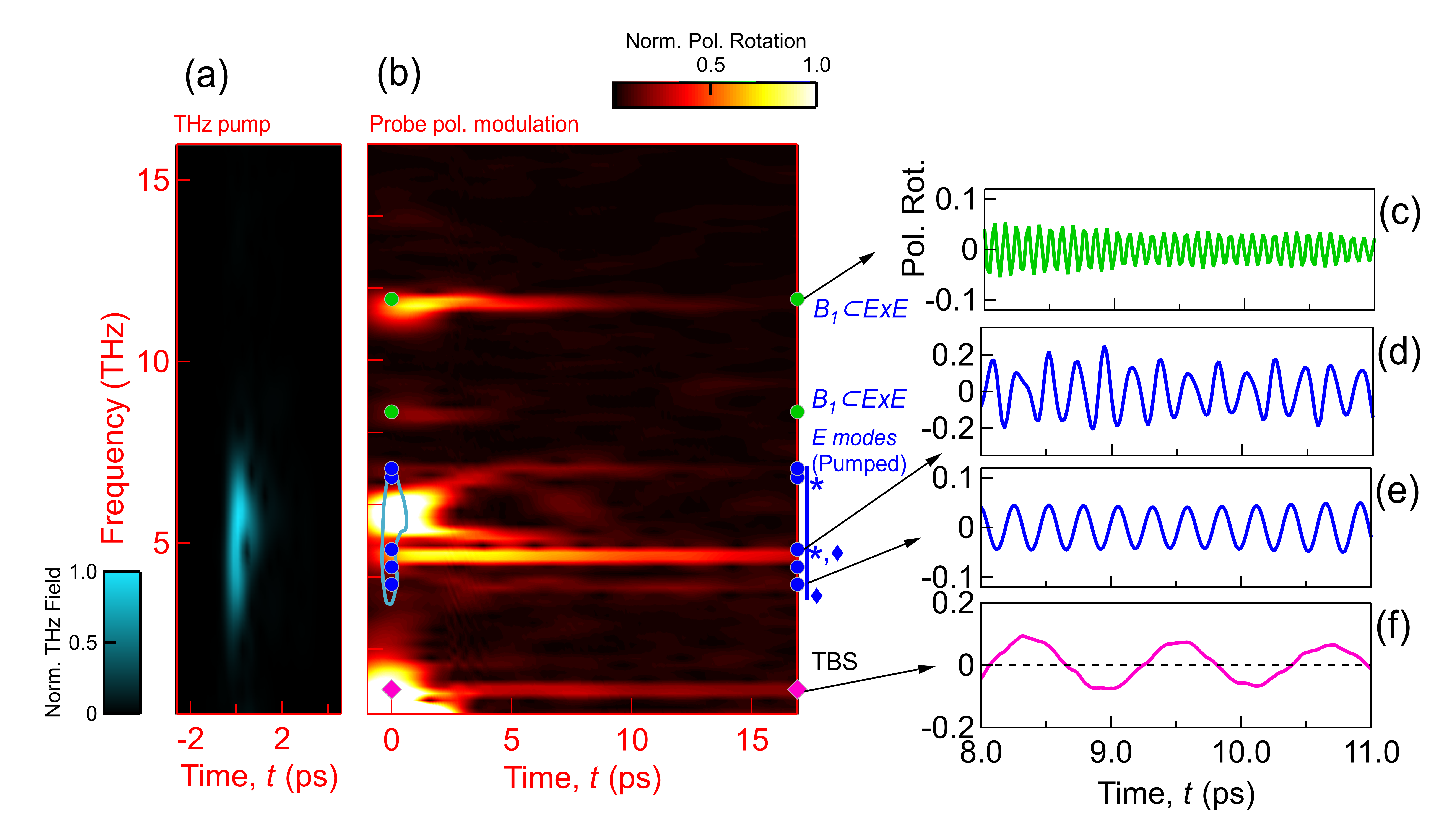}
\caption{{\bf Time-frequency THz-driven dynamics.} (a) THz pump electric-field 
amplitude. (b) Pump-induced polarization modulation in SrCu$_2$(BO$_3$)$_2$ at 
4 K. In panel (a), the color contours around $t = 0$ are based on the FWHM 
isoline of the amplitude. In panel (b), the blue circles indicate the 
frequencies of the $E$-symmetric IR-active phonon modes that are pumped 
directly, where our measured values of 3.80, 4.27, 6.75, and 7.00 THz match 
within the experimental error with those of Ref.~\cite{Homes09}, whereas the 
4.60 THz mode we measure is tabulated there as 4.75 THz. Green circles 
show the frequencies of $B_1$-symmetric modes at 8.57 THz and 11.7 THz 
\cite{Homes09} that are not IR-active but are excited by sum-frequency 
processes, which are expected to involve respectively the $E$-symmetric modes 
indicated with diamonds (3.80 THz and 4.60 THz) and with asterisks (4.60 THz 
and 7.00 THz). The pink diamond marks the frequency of the TBS, $\omega_{\rm 
TBS} = 0.87$ THz (taken from Refs.~\cite{Lemmens00,Gozar05}); this mode is 
not electromagnetically active and is driven by the nonlinear 
(difference-frequency) spin-phonon coupling discussed in Sec.~IV. 
(c-f) Temporal dynamics, obtained by numerical band-pass filters, of the 
normalized polarization rotations measured for some of the primary excited 
modes.} 
\label{uqmfs2}
\end{figure*}

The THz-pump and optical-probe set-up is shown in Fig.~\ref{uqmfs1}(a) 
\cite{Vicario20}. The output of a 20 mJ, 55 fs, 800 nm Ti:sapphire laser 
was used to drive an optical parametric amplifier (OPA), which provides 
ultrashort, multi-mJ pulses \cite{Giorgianni19}. Single-cycle THz pulses 
were generated by optical rectification, using a crystal of DAST 
(4-N,N-dimethylamino-4'-N'-methyl-stilbazolium tosylate, from Rainbow 
Photonics), of the OPA signal at 1.5 $\mu$m. The OPA pulse energy of 
3.2 mJ gave an NIR pump fluence at the crystal surface of approximately 
5 mJcm$^{-2}$. Three low-pass filters, two with a 20 THz cut-off frequency 
and one with 10 THz, were used after the THz generation step to block the 
residual OPA beam, and gave an extinction ratio for the pump in excess of 
105. For the pump pulses, an additional high-pass filter with a cut-off of 
4.2 THz was used to drive the primary IR-active phonons while ensuring a 
negligible spectral weight at the frequency of the leading magnetic modes 
[Fig.~\ref{uqmf3}(a)]. To select the four different pump pulses shown in 
Fig.~\ref{uqmf4}(c), we used respectively a 2 THz low-pass filter, a 3 
THz band-pass filter, a 4.2 THz high-pass filter coupled with a 6 THz 
low-pass filter, and a 6 THz band-pass filter. 

For the measurement of properties dependent on the pump strength, the THz 
electric field was tuned by three wire-grid polarizers. Peak electric fields 
were reached by tight focusing of the THz beam using three parabolic mirrors 
\cite{Giorgianni19}. To estimate the electric-field strength of the pump 
pulses shown in Fig.~\ref{uqmf1}, we apply the formula \cite{Thomson07}
\begin{equation}
E_{\rm THz} = \sqrt{\frac{z_0 E_p 4 \sqrt{\ln 2}}{\pi \sqrt{\pi} w^2 \tau_{\rm 
FWHM}}}, 
\label{eethz}
\end{equation}
where $z_0$ is the vacuum impedance and we measured (i) the THz energy 
per pulse, $E_p = 0.8$ $\mu$J, using a calibrated THz energymeter (Gentec 
THZ12D3S-VP-D0); (ii) the beam waist, $w = 92$ $\mu$m, obtained by a 
Gaussian fit of the beam profile [Fig.~\ref{uqmfs1}(b)], which was measured 
with a micro-bolometric THz camera (NEC IRV-T0830); (iii) the pulse duration, 
$\tau_{\rm FWHM} = 0.21$ ps, obtained from the FWHM of the Gaussian envelope 
fitting the temporal intensity waveform of the THz pump [Fig.~\ref{uqmfs1}(c)], 
which was taken as the square of the electric field measured at the sample 
position by electro-optic sampling in a 200 $\mu$m-thick (110) GaP crystal 
with a 50 fs, 800 nm gating pulse obtained as a fraction of the Ti:sapphire 
beam. Our estimated electric-field strength, $E_{\rm THz} = 3.2$ MVcm$^{-1}$, 
is similar to other values in recent literature \cite{Liu17,Agranat18}.

The gating pulse was used to probe the ultrafast pump-induced polarization 
dynamics of the sample, which were measured by splitting the probe beam into 
two orthogonal components with a Wollaston prism. The THz electric field of 
the pump was polarized in the sample plane along the direction perpendicular 
to the optical table. The polarization of the probe relative to the sample 
was offset by 45$^\circ$ from the pump polarization. All measurements were 
performed in a He cryostat, which allowed a minimum sample temperature of 
3.5 K to be reached. 

For a quantitative analysis of the effect of $E_{\rm THz}$ on the sample, the 
peak polarization, $P_m$, induced by the THz pulse resonant with a generic 
phonon mode, $m$, is \cite{Mitrano16}
\begin{equation}
P_m = \frac{\sigma_1 (\omega_m)}{\omega_m} \tilde{E}_{\rm THz},
\label{epm}
\end{equation}
where $\omega_m$ is the angular frequency and $\sigma_1 (\omega_m)$ the optical 
conductivity of the driven phonon. The polarization arises from the modulation 
of the dipole moment of the crystal, $P_m = n_d \delta_m \mu_m$, where $\mu_m
 = e |\sum_{im} \hat{u}_{im} Z_{\rm eff}^i|$ is the magnitude of the charge 
displacement due to phonon mode $m$, $\hat{u}_{im}$ is the normalized vector 
of displacements of each atom, $i$, in phonon mode $m$, $Z_{\rm eff}^i$ are the 
Born effective charges of each atom, $n_d$ is the number of dipoles per unit 
volume, and $\delta_m$ is the maximum displacement coordinate of the phonon. 
The driving field, ${\tilde E}_{\rm THz} = \beta_m E_{\rm THz}$ in Eq.~(\ref{epm}), 
is the effective electric field inside the sample acting on the phonon 
\cite{Thomson07}, where $\beta_m = 1 - R_m$ is determined by the reflectivity 
at frequency $\omega_m$.

\begin{figure*}[t]
\includegraphics[width=0.75\linewidth]{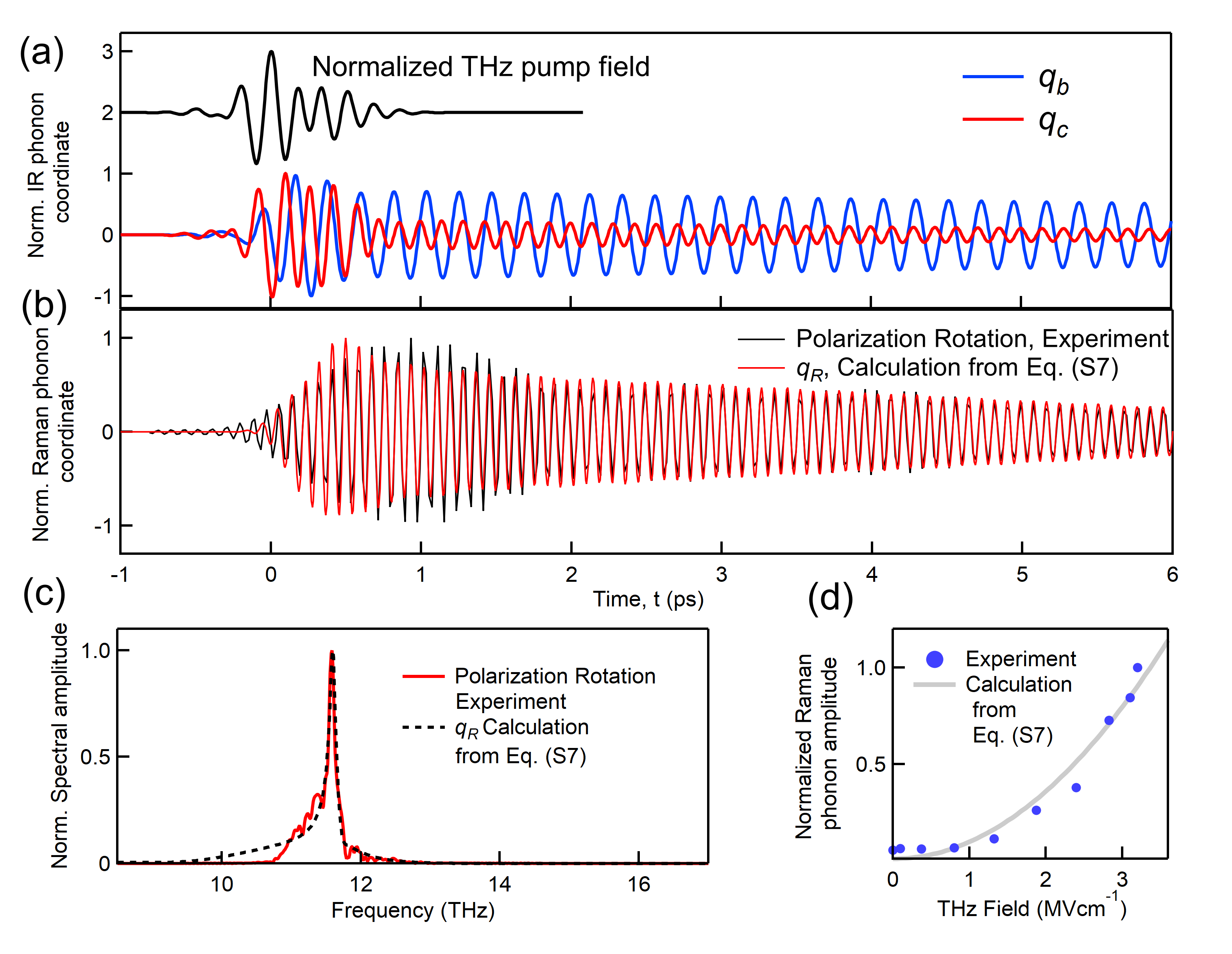}
\caption{{\bf Sum-frequency ionic Raman excitation: experimental evidence 
and model.} (a) Normalized experimental THz driving field (black) and 
normalized time-dependent THz-driven atomic displacements, calculated from 
Eqs.~(\ref{ecpdeb}) and (\ref{ecpdec}), of the IR-active phonon modes $q_b$ 
at $\omega_b = 4.60$ THz (blue) and $q_c$ at $\omega_c = 7.0$ THz (red). (b) 
Time-dependent phonon-driven atomic displacement, $q_R$, of the Raman-active 
phonon mode, obtained from Eq.~(\ref{ecpder}) and compared to the normalized 
polarization rotation (Fig.~\ref{uqmfs2}) measured by using a numerical 
band-pass filter to remove the components of the lower-frequency modes. (c) 
Fourier-transformed amplitude of $q_R$ and of the polarization rotation shown 
in panel (b). (d) Measured amplitude of $q_R$ compared with the quadratic 
electric-field dependence given by Eq.~(\ref{ecpder}). }
\label{uqmfs3}
\end{figure*}

To illustrate the estimation of the $\delta_m$ values induced by the THz 
electric field, we take the example of the 4.60 THz phonon mode, which is 
the strongest single feature of the driven response (Fig.~\ref{uqmf3}) and 
is labelled $m = b$ in Sec.~IV. Thus we use 
$\omega_b = 2 \pi 4.6 \times 10^{12}$ s$^{-1}$, $\sigma_1 (\omega_b) = 137$ 
$\Omega^{-1}$ cm$^{-1}$ from Ref.~\cite{Homes09}, $n_d = V^{-1}$ with $V = 5.71 
\times 10^{-22}$ cm$^{-3}$ the volume of the unit cell, and the value $\mu_b = 
0.6 e$ taken from our DFT calculations (below). The bandwidth of the external 
THz field, approximately 2.7 THz FWHM [Fig.~\ref{uqmf3}(a)], is large compared 
to the linewidth of the phonon mode [0.1 THz FWHM, Fig.~\ref{uqmf4}(c)], and 
by accounting for the reflection of the external field [Fig.~\ref{uqmf4}(c)] 
we estimate $\beta_b \simeq 0.02$. Thus we deduce a maximum displacement of 
$\delta_b = 0.17$ \AA, which is comparable to that estimated in SrTiO$_3$ 
\cite{Kozina19}. 

\section{Time-frequency analysis of TH\lowercase{z}-driven dynamics and mode 
symmetries}
\label{stfa}

Figure \ref{uqmfs2} reports the temporal profile and frequency content 
measured for the THz pump pulse [Fig.~\ref{uqmfs2}(a)] and the resulting 
polarization rotations induced in the SrCu$_2$(BO$_3$)$_2$ sample at 4 K 
[Fig.~\ref{uqmfs2}(b)]. The spectral decomposition was computed using a 
Hamming sliding-window fast Fourier transform. It is clear that the 
distribution of frequencies in the THz pump pulse [Fig.~\ref{uqmfs2}(a)] 
causes not only a direct resonant driving of $E$-symmetric, IR-active 
phonons but also a less direct, nonlinear driving of $B_1$-symmetric, 
Raman-active phonons at higher frequencies [Figs.~\ref{uqmfs2}(b-c)]; for 
this the low-temperature point group, $D_{2d}$ (space group I$\bar{4}$2m), 
permits the excitation of $B_1$ $(\subset E \times E)$ phonons by the 
composition of two $E$-symmetric phonons. Similarly, the magnetic excitation 
from the singlet ground state to the TBS mode can be interpreted as a resonance 
driven by the difference-frequency harmonic components of the $E$-symmetric 
phonons shown in Figs.~\ref{uqmfs2}(d-e). The TBS [Fig.~\ref{uqmfs2}(f)]
also has $B_1$ symmetry, as determined by Raman spectroscopy \cite{Lemmens00, 
Gozar05}, indicating that its driving relies on the same symmetry composition 
($B_1 \subset E \times E$). We comment that, because both the singlet ground 
state and the TBS are nonmagnetic ($S = 0$), the origin of the polarization 
rotation measured in SrCu$_2$(BO$_3$)$_2$ must lie in lattice (birefringence) 
effects, which are clearly enhanced by the spin-lattice coupling.

\begin{figure*}[t]
\includegraphics[width=0.85\linewidth]{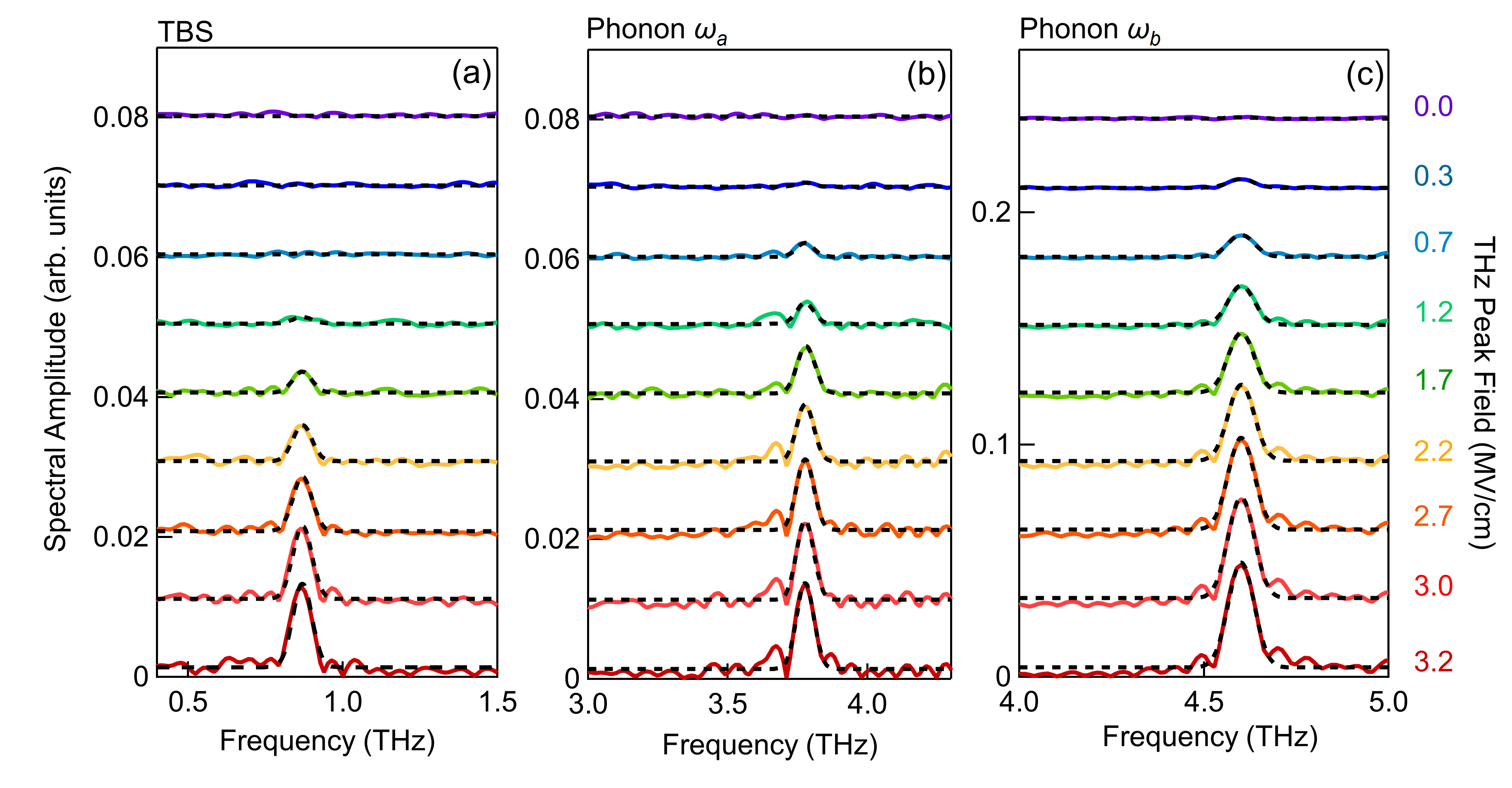}
\caption{{\bf Extraction of mode amplitudes.} Spectral amplitude computed for 
the interval $1.9 \leq t \leq 14.4$ ps and shown for a range of THz pump field 
strengths with an added vertical offset. (a) TBS. (b) Phonon $\omega_a$. (c) 
Phonon $\omega_b$. Short-dashed black lines show fits to a Gaussian function, 
$\mathcal{G}(a, \omega_0, w)$, with the amplitude $a$ as a free parameter. The 
center frequency ($\omega_0$) and Gaussian width (standard deviation, $w$) 
parameters for the TBS are $\omega_0 = 0.87$ THz and $w = 0.06$ THz, for phonon 
mode $a$ are $\omega_0 = 3.77$ THz and $w = 0.06$ THz, and for phonon mode $b$ 
are $\omega_0 = 4.60$ THz and $w = 0.07$ THz.}
\label{uqmfs5}
\end{figure*}

\section{TH\lowercase{z}-driven nonlinear phonon dynamics}
\label{snpd}

Our results (Sec.~III) demonstrate that the resonant dynamic distortion 
of the lattice in response to intense and coherent THz excitation creates 
nonlinear channels for the transfer of energy to both magnetic and phononic 
modes. The phenomenon of sum-frequency ionic Raman scattering has been 
investigated recently in both experiment \cite{Foerst11,Melnikov18} and 
theory \cite{Juraschek18}, and our results include its clearest observation 
to date. To describe the driven nonlinear lattice dynamics that we observe, 
we consider two IR-active phonons with normal coordinates $q_b$ and $q_c$, 
with corresponding frequencies $\omega_b$ and $\omega_c$, and a Raman-active 
phonon with coordinate $q_R$ and frequency $\omega_R$. By discarding terms 
quadratic in $q_R$, on the assumption that the amplitude of the Raman mode 
will be much smaller than that of the IR modes driven by the THz pump 
($|q_{b,c}| \gg |q_R|$), the minimal lattice potential to cubic order 
(i.e.~lowest anharmonic order) is
\begin{eqnarray} 
V(q_b,q_c,q_R) & = & {\textstyle \frac12} \omega_b^2 q_b^2 + {\textstyle \frac12}
\omega_c^2 q_c^2 + {\textstyle \frac12} \omega_R^2 q_R^2 \\ 
& & \;\;\;\; + [c_{bb,R} q_b^2 + c_{bc,R} q_b q_c + c_{cc,R} q_c^2] q_R, \nonumber
\label{eahi}
\end{eqnarray} 
where the $c$ coefficients specify the leading nonlinear coupling terms 
between the IR and Raman phonons. The equation of motion for a generic 
THz-driven, IR-active phonon mode, $m$, takes the form
\begin{equation}
{\ddot q}_m + \gamma_m {\dot q}_m = - \frac{\partial [V - B_m q_m E_{\rm THz} 
(t)]}{\partial q_m},
\label{epeom}
\end{equation}
with $\gamma_m$ the damping rate and $B_m$ the dipole coupling constant 
introduced in Eq.~(5). For a phonon that is Raman-active but not IR-active, 
the driving term is only $\partial V/\partial q_R$. The coherent 
THz-driven lattice dynamics are then described in the time domain from 
Eq.~(\ref{eahi}), to leading order in $q_R$, by the coupled differential 
equations
\begin{eqnarray} 
\label{ecpdeb} \!\!\!\!\!\!\!\! {\ddot q}_b + \gamma_b {\dot q}_b + \omega_b^2 
q_b & = & - B_b E_{\rm THz} (t), \\
\label{ecpdec} \!\!\!\!\!\!\!\! {\ddot q}_c + \gamma_c {\dot q}_c + \omega_c^2 
q_c & = & - B_c E_{\rm THz} (t), \\
\label{ecpder} \!\!\!\!\!\!\!\! {\ddot q}_R \! + \! \gamma_R {\dot q}_R \! +
 \! \omega_R^2 q_R \! & = & \! - [c_{bb,R} q_b^2 \! + \! c_{bc,R} q_b q_c \! + \! 
c_{cc,R} q_c^2].
\end{eqnarray} 
Clearly Eq.~(\ref{ecpder}) for the Raman phonon describes a damped harmonic 
oscillator driven by terms quadratic in the IR-active phonon displacements, 
resulting in sum-frequency excitation processes. The effectiveness of this 
driving then depends on the proximity of the combinations $2 \omega_b$, 
$2 \omega_c$, and $\omega_b + \omega_c$ (and indeed $\omega_c - \omega_b$) 
to $\omega_R$.

The clearest example in Fig.~\ref{uqmfs2}(b) is the $B_1$ Raman mode 
with frequency $\omega_R  = 11.7$ THz and damping $\gamma_R = 0.2$ THz 
[Fig.~\ref{uqmfs2}(c)]. The two THz-pumped $E$-symmetric phonons marked 
by the asterisks have parameters $\omega_b = 4.60$ THz, $\gamma_b = 0.03$ 
THz and $\omega_c = 7.00$ THz, $\gamma_c = 0.10$ THz. Figure \ref{uqmfs3}(a) 
shows the temporal evolution of $q_b$ and $q_c$ calculated from 
Eqs.~(\ref{ecpdeb}-\ref{ecpder}) using the experimental THz pump field. As 
expected from the respective phonon frequencies, the Raman coordinate, $q_R$, 
is dominated by coherent oscillations at frequency $\omega_b + \omega_c$ 
[Fig.~\ref{uqmfs3}(b)], confirming the sum-frequency ionic Raman mechanism. 
The phenomenon is equally clear in the frequency domain, where the Fourier 
amplitudes of the measured and calculated dynamics are compared in 
Fig.~\ref{uqmfs3}(c). Because the amplitude of $q_R$ varies linearly with 
each individual amplitude $q_b$ or $q_c$ [Eq.~(\ref{ecpder})], it should 
scale quadratically with the THz field strength, as Fig.~\ref{uqmfs3}(d) 
confirms. We note again that all the phonon frequencies used in our analysis 
are the experimental values obtained from our pump-probe measurement, which 
with one exception match the FTIR spectroscopy measurements of 
Ref.~\cite{Homes09} to within the experimental uncertainties; as a result 
of this match, we took all the phonon damping parameters ($\gamma_m$) from 
Ref.~\cite{Homes09}. The exception was the phonon frequency $\omega_b = 4.60$ 
THz, found at 4.75 THz by FTIR spectroscopy, where our measured value gave a 
significantly better account of the $B_1$ Raman frequency we observed at 11.58 
THz. A similar analysis can be performed for the $B_1$ phonon mode at 8.57 
THz, with driving by $\omega_b$ combined with the IR-active phonon at 
$\omega_a = 3.80$ THz [both marked by blue diamonds in Fig.~\ref{uqmfs2}(b)].

\section{Fitting of phonon and TBS modes}

The amplitudes of the $\omega_a$ and $\omega_b$ phonon modes shown in 
Fig.~\ref{uqmf4}(a), and of the TBS mode shown in Fig.~\ref{uqmf4}(b), were 
determined together with their uncertainties by a Gaussian fitting procedure 
in the frequency domain. Figure \ref{uqmfs5} shows the spectral peaks of the 
TBS mode, and of the two phonon modes, for a range of values of the peak 
electric field of the THz pump pulse, with the Gaussian fits shown as dashed 
black lines. In fitting a Gaussian function, $\mathcal{G}(a, \omega_0, w)$, 
to these peaks, the only free parameter was the peak amplitude, $a$, whereas 
the center frequency, $\omega_0$, and the Gaussian width, $w$, were fixed for 
every peak. The amplitudes and their uncertainties were then determined by the 
least-squares Gaussian fit method \cite{Sivia06}.

\begin{figure}[t]
\includegraphics[width=\columnwidth]{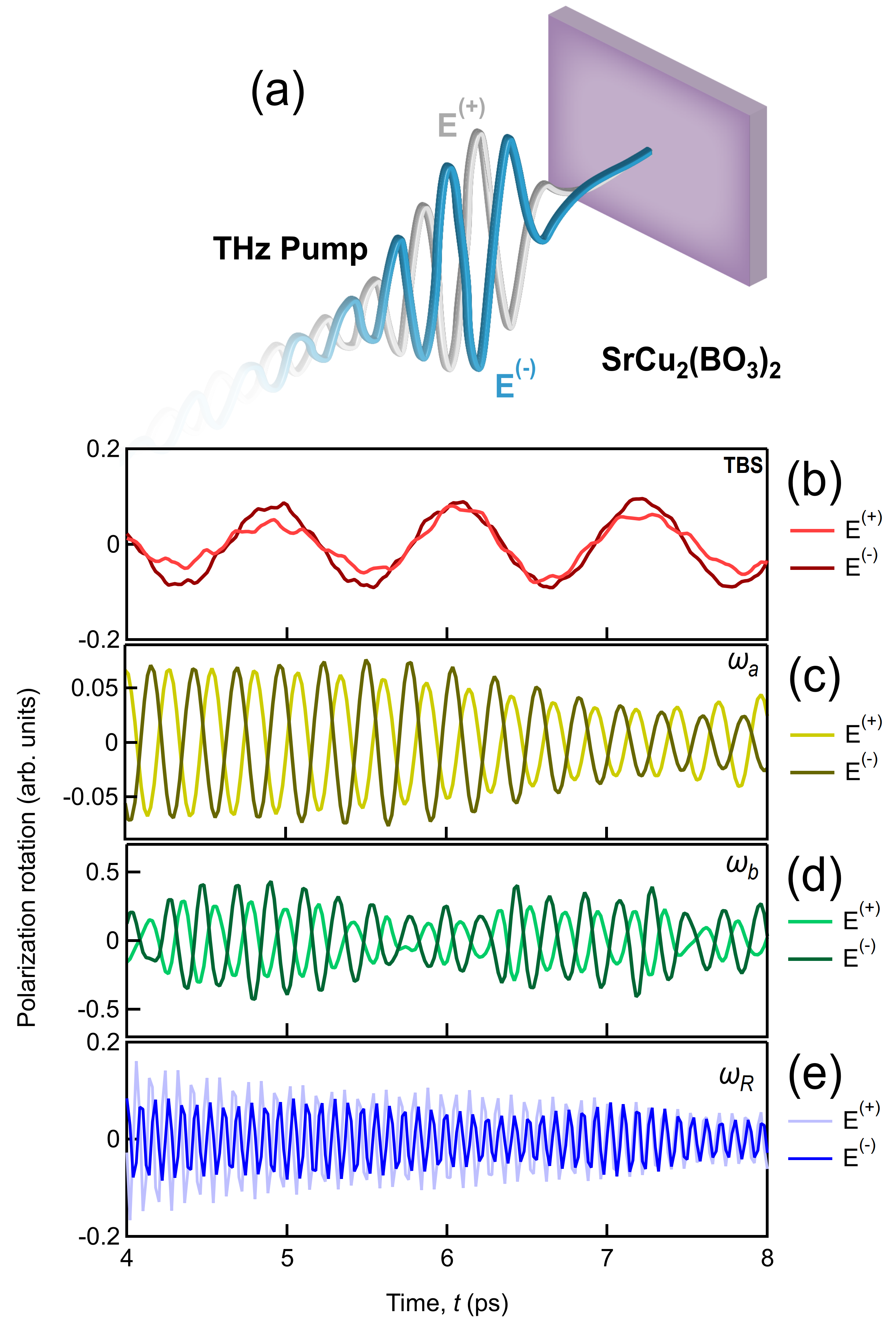}
\caption{{\bf Pump polarity inversion.} (a) Representation of two THz 
pump pulses with inverted electric-field polarities, $E^{(+)}$ and $E^{(-)}$. 
(b-e) Comparison of the dynamics of the magnetic and lattice modes for these 
two pump polarities.} 
\label{uqmfs4}
\end{figure}

The nonlinear dependence of the TBS amplitude on the THz electric-field 
strength can be also be read from Fig.~\ref{uqmfs5}(a), by comparing it with 
the linear dependence of $\omega_a$ in Fig.~\ref{uqmfs5}(b). Both modes have 
comparable amplitudes at the maximum THz electric field, making clear the 
much faster decrease in the TBS amplitude at lower field strengths. 
 
\section{Dependence on polarity of the TH\lowercase{z} pump field}

As Fig.~\ref{uqmf4}(b) shows, the amplitude of the phonon-driven TBS 
excitation is proportional to the square of the driving electric field of 
the THz pump pulse. A further important test of our observations and modelling 
is therefore to invert the polarity of this field, as represented in 
Fig.~\ref{uqmfs4}(a). To implement this sign-inversion experimentally, we 
rotate the crystal for THz generation by 180$^\circ$ and compare the driven 
dynamics of the polarization-rotation signals. The results in the time domain 
are shown in Figs.~\ref{uqmfs4}(b-e). Because the $E$-symmetric (IR-active) 
phonon modes are excited directly, the carrier-envelope phase of the THz pump 
is imprinted onto them and Figs.~\ref{uqmfs4}(c-d) confirm a phase shift of 
180$^\circ$. 

By contrast, the dynamics of the TBS [Fig.~\ref{uqmfs4}(b)] are invariant 
on changing the electric-field polarity, fully consistent with a quadratic 
driving mechanism. The dynamics of the Raman-active phonon at 11.6 THz, 
presented in App.~C, are similarly insensitive to the change of polarity 
[Fig.~\ref{uqmfs4}(e)], as expected of a sum-frequency process. We note that 
perfect inversion of the IR phonon signals, and perfect overlap of the TBS 
and Raman phonon signals, are actually obtained for a time shift of $-20$ fs 
in the $E^{(-)}$ signal. This can be attributed to the fact that rotation of 
the THz generation crystal may produce a small time-delay in the event of 
an inhomogeneity in its thickness (20 fs corresponds to approximately 6 
$\mu$m in vacuum).

\begin{figure}
\includegraphics[width=\columnwidth]{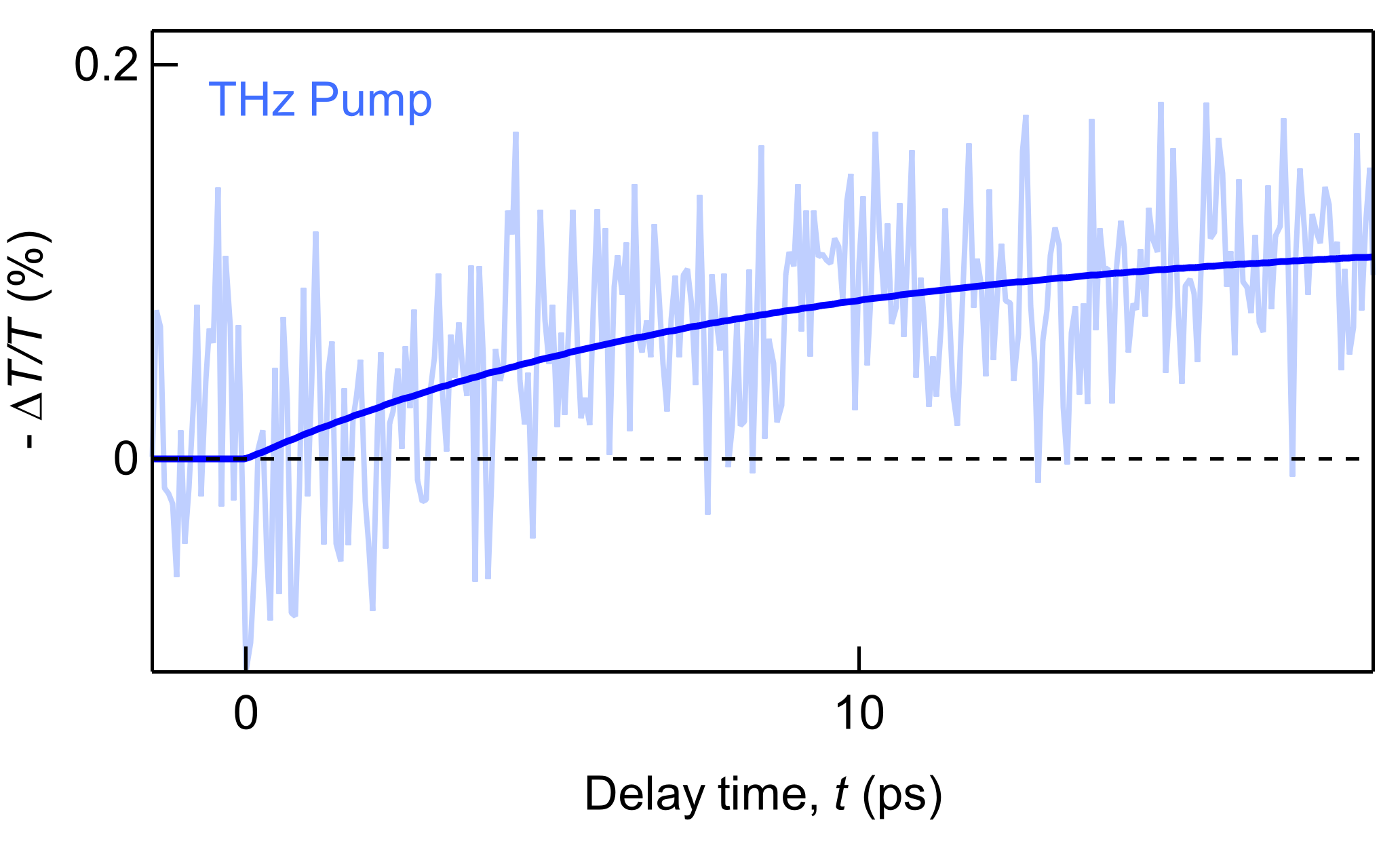}
\caption{{\bf Transmission modulation.} NIR modulation of the THz transmission 
measured in SrCu$_2$(BO$_3$)$_2$ under the same experimental conditions as the 
NIR polarization rotation shown in Fig.~\ref{uqmf1}(c). The blue solid 
line is a fit to the functional form $1 - e^{-t/\tau}$ with $\tau = 8.3$ ps.}
\label{uqmfs6}
\end{figure}

\section{TH\lowercase{z}-pumped NIR transmission modulation}

To demonstrate that ultrafast electronic processes, such as the displacive 
excitation of phonons \cite{Zeiger92}, are not involved in the THz generation 
mechanism of the coherent phonons and TBS modes we observe, we have performed 
THz-pump, NIR-probe spectroscopy to measure the transient transmission 
modulation, $\Delta T/T$. Working under the same experimental conditions as 
for the measurements in Fig.~\ref{uqmf1}(c), we measured the THz-induced 
NIR $\Delta T/T$ at a photon energy of 1.55 eV and show the results as a 
function of the pump-probe time delay in Fig.~\ref{uqmfs6}. The transient 
transmission can monitor the presence of ultrafast THz-induced carrier 
generation, which would appear as a peak at very short times, but the observed 
dynamics show no signature of such processes. Instead the $\Delta T/T$ dynamics 
show only a slow decrease of the sample transmission upon THz irradiation. 

In solids with impurity states or in narrow-gap insulating systems, an intense 
THz pulse can generate carriers through impact ionization or tunneling 
processes \cite{Kampfrath13,Hubmann20}, whose dynamics are confined to 
the duration of the THz pump pulse (i.e.~a few picoseconds). Carrier excitation 
can generate vibrational or spin modes through a displacive process if the 
excitation dynamics are shorter than half an oscillation period of the mode 
\cite{Giorgianni22,Jnawali21}. However, the measured $\Delta T/T$ is not 
compatible with this scenario: no free carriers are excited and therefore it is 
self-evident that displacive excitations play no role in the phenomenon we have 
observed. Taken together with the evidence presented in Sec.~III, this result 
further reinforces the magnetophononic mechanism.

\begin{figure*}[t]
\includegraphics[width=0.94\linewidth]{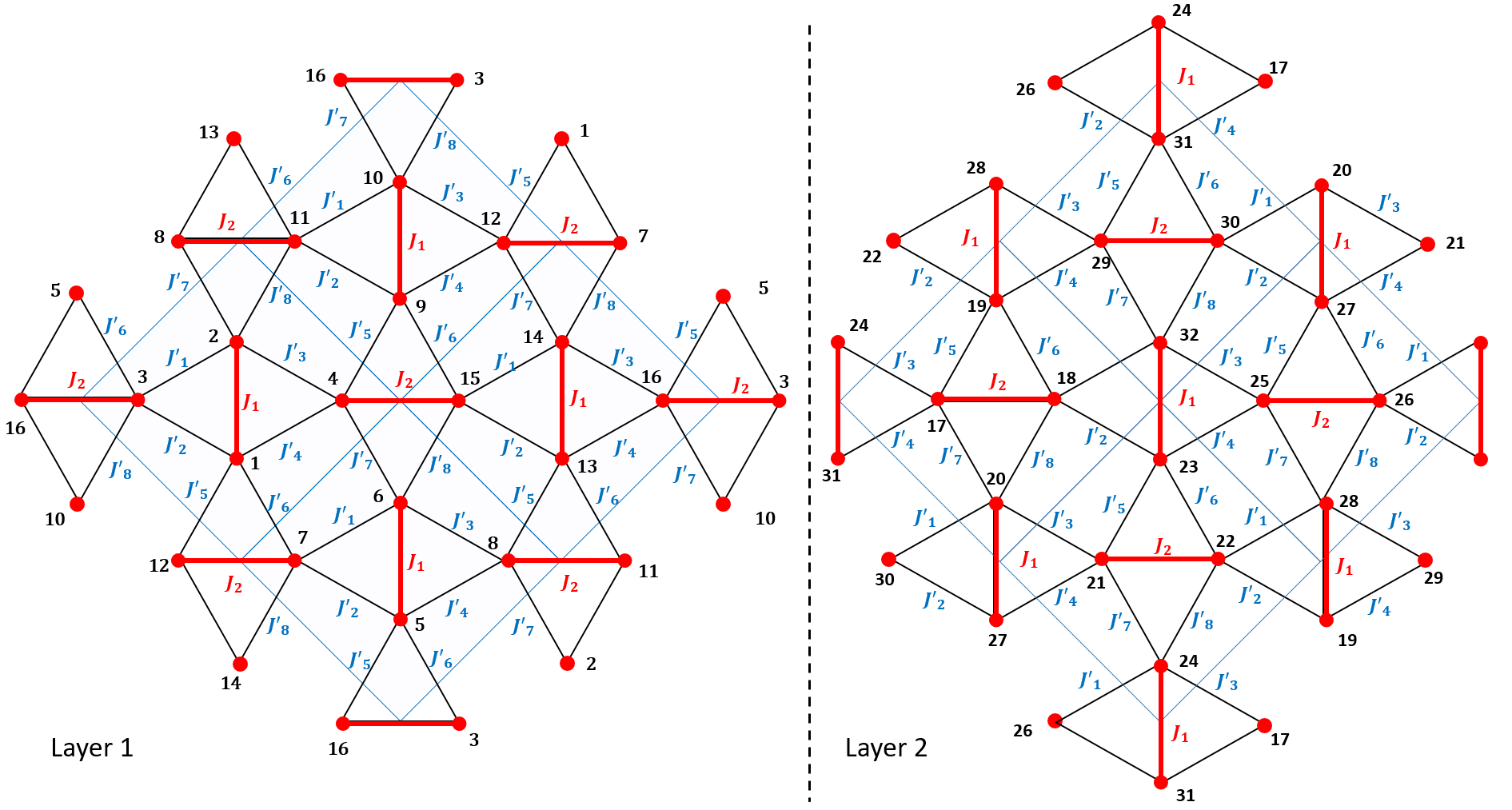}
\caption{{\bf Frozen-phonon density-functional theory.} Symmetry-related 
magnetic interaction parameters in the spin network obtained for an arbitrary 
``frozen'' configuration of the phonons in SrCu$_2$(BO$_3$)$_2$. Red dots 
represent the $S = 1/2$ spins at the Cu$^{2+}$ sites in SrCu$_2$(BO$_3$)$_2$. 
For every normal mode of the lattice, the system has two different values of 
the intradimer interaction, $J_1$ and $J_2$ (red lines), and eight different 
interdimer interaction terms, $J_1'$, $J_2'$, \dots, $J_8'$ (black). Shown are 
the 2$\times$2 supercells in each of the two layers of the structural unit 
cell (a total of 32 magnetic sites) required to obtain enough independent 
spin configurations to determine all 11 unknown parameters.}
\label{uqmfs7}
\end{figure*}

The $\Delta T/T$ dynamics shown in Fig.~\ref{uqmfs6} can be fitted by a single 
exponential with a time constant of $\tau = 8.3$ ps (solid blue line), and thus 
can be ascribed to a slight increase in temperature arising from the relaxation 
of lattice phonons. One may estimate the maximum increase in temperature due to 
THz irradiation as being less than 1 K, based on the low-temperature specific 
heat of 5 J/mol/K \cite{Vasil_ev01}. The decrease of the NIR transmission 
with increasing temperature is consistent with the fact that the steady-state 
NIR transmission decreases by approximately 50\% upon heating from 4 K to 300 K.

\section{Phonon modulation of magnetic interactions}
\label{sdft}

We have performed a hierarchy of DFT calculations in order to obtain 
quantitative estimates of the effects of the driven IR phonons on the 
magnetic system in SrCu$_2$(BO$_3$)$_2$. 

\subsection{DFT calculations at equilibrium}
 
First we used Quantum Espresso \cite{Giannozzi17} to compute the total 
lattice and magnetic energies of SrCu$_2$(BO$_3$)$_2$ at equilibrium. 
For these calculations we worked in the structural unit cell of the 
system, which is tetragonal and contains 44 atoms (of which 8 are Cu atoms) 
in two ``Shastry-Sutherland'' layers. The electron-ion interactions were 
modelled using pseudopotentials from the curated SSSP library \cite{Prandini18,
Prandini20}. The plane-wave cut-off was set at 750 eV (6000 eV for the charge 
density) and for sampling of the Brillouin zone we used a 6$\times$6$\times$6 
$k$-point grid. Structural relaxation was continued until each component of 
the force acting at every atom was less than 0.0025 eV/\AA~and the pressure 
(defined as ${\textstyle \frac{1}{3}} Tr[\overleftrightarrow{\sigma}]$, with 
$\overleftrightarrow{\sigma}$ being the stress tensor) was below 0.5 kbar. 
These parameters ensure errors smaller than 1 meV/atom within the total energy 
and, more importantly, smaller than 1 K on the magnetic interactions (which 
depend only on total-energy differences, and thus converge faster and better 
than do the total values). By comparing the energies of different spin 
configurations computed with the same relaxed structure, we fixed the value 
of the effective Hubbard term to $U = 11.4$ eV by reproducing the interaction 
parameters $J = 84.0$ K and $J' = 49.7$ K \cite{Radtke08}. The lattice 
parameters we obtain for this value of $U$, the $T = 0$ DFT + $U$ structure, 
agree with the measured low-temperature structure of SrCu$_2$(BO$_3$)$_2$ 
\cite{Vecchini09} to within 1\% for the $a$ and $b$ axes and 3\% for the 
$c$ axis. The Born effective charges, $Z_{\rm eff}^i$, used in Sec.~III to 
estimate the maximum displacements, $\delta_m$, were computed for the system 
in the AFM configuration using the PHONON module of Quantum Espresso and the 
PBE functional \cite{Perdew96}.

\subsection{DFT calculations for frozen phonons}
 
In order to extend our DFT calculations to include the nonequilibrium atomic 
configurations in the presence of lattice excitations, we first made a phonon 
symmetry analysis of the different interatomic paths. From this we deduced 
that the most general magnetic state in the presence of a phonon distortion 
$q_t$ is characterized by ten different interaction parameters in the unit 
cell, two values of $J (q_t)$ and eight of $J' (q_t)$, as shown in 
Fig.~\ref{uqmfs7}. To compute this number of unknown parameters, it is 
necessary to work on a 2$\times$2$\times$1 supercell, meaning to use four 
of the basic magnetic unit cells of SrCu$_2$(BO$_3$)$_2$. In these supercell 
calculations we reduced the $k$-point grid to 2$\times$2$\times$3, and 
verified that this sampling density provided essentially equivalent results. 

\begin{table*}[t]
\caption{\label{t1} Eleven independent spin configurations used for one 
determination of the magnetic interaction parameters. Columns from 1 to 16 
are the magnetic sites shown in Layer 1 of Fig.~\ref{uqmfs7}. 1 denotes an 
up-spin ($S_z = 1/2$) and $-1$ a down-spin ($S_z = - 1/2$).}
\centering
\begin{tabular}{c|cccccccccccccccc}
Config. & \multicolumn{16}{c}{Magnetic site} \\
\# & 1 & 2 & 3 & 4 & 5 & 6 & 7 & 8 & 9 & 10 & 11 & 12 & 13 & 14 & 15 & 16 \\
\hline
1 & $\; -1 \;$ & $\; -1 \;$ & 1 & 1 & $\; -1 \;$ & $\; -1 \;$ & 1 & 1 & 1 & 1
 & $\; -1 \;$ & $\; -1 \;$ & 1 & 1 & $\; -1 \;$ & $\; -1 \;$ \\
2 & $-1$ & 1 & $\; -1 \;$ & $\; -1 \;$ & 1 & 1 & $\; -1 \;$ & $\; -1 \;$
 & $\; -1 \;$ & $\; -1 \;$ & 1 & 1 & $\; -1 \;$ & $\; -1 \;$ & 1 & 1 \\
3 & 1 & 1 & $-1$ & $-1$ & 1 & 1 & $-1$ & $-1$ & 1 & $-1$ & 1 & 1 & $-1$ & $-1$
 & 1 & 1 \\
4 & $-1$ & 1 & $-1$ & $-1$ & 1 & $-1$ & $-1$ & 1 & 1 & $-1$ & $-1$ & 1 & $-1$
 & 1 & 1 & 1 \\
5 & $-1$ & $-1$ & 1 & 1 & $-1$ & $-1$ & 1 & 1 & $-1$ & $-1$ & 1 & 1 & $-1$
 & $-1$ & 1 & 1 \\
6 & $-1$ & $-1$ & $-1$ & 1 & $-1$ & $-1$ & 1 & 1 & $-1$ & $-1$ & 1 & 1 & $-1$	
 & $-1$ & 1 & 1 \\
7 & $-1$ & $-1$ & 1 & 1 & $-1$ & $-1$ & 1 & 1 & $-1$ & $-1$ & 1 & 1 & $-1$
 & 1 & 1 & 1 \\
8 & $-1$ & $-1$ & 1 & 1 & $-1$ & $-1$ & 1 & 1 & $-1$ & $-1$ & 1 & 1 & $-1$
 & $-1$ & 1 & $-1$ \\
9 & $-1$ & $-1$ & 1 & 1 & $-1$ & $-1$ & 1 & 1 & $-1$ & 1 & $-1$ & $-1$ & 1	
 & $-1$	& $-1$ & 1 \\
10 & $-1$ & $-1$ & 1 & 1 & $-1$ & $-1$ & 1 & 1 & $-1$ & 1 & 1 & $-1$ & 1 & $-1$	
 & $-1$ & $-1$ \\
11 & $-1$ & $-1$ & 1 & 1 & $-1$ & $-1$ & 1 & 1 & 1 & $-1$ & 1 & $-1$ & $-1$
 & 1 & $-1$ & $-1$
\end{tabular}
\end{table*}

We performed total-energy calculations for the lattice structures obtained by 
systematic displacement of all atoms according to the normal coordinates of 
the strongest two phonon modes measured in experiment [Fig.~\ref{uqmfs2}(a)], 
$q_a$ at $\omega_a = 3.80$ THz and $q_b$ at $\omega_b = 4.60$ THz. While it is 
possible to include many more of the phonons observed in App.~B, here 
we have focused rather on optimizing our treatment of the time and frequency 
domains. For each distorted structure, we computed the electronic ground state 
as a function of the displacement amplitudes of the two phonons and for 11 
different configurations of up- ($S_z = 1/2$) and down-oriented spins ($S_z =
 - 1/2$), from which we obtained 11 linearly independent equations in order 
to determine all the interaction parameters in the system. This process is 
represented in Table I for one set of 11 spin configurations in Layer 1 of 
Fig.~\ref{uqmfs7} and the corresponding equations are

\onecolumngrid

\begin{widetext}
\begin{eqnarray}
E(1) & = & E_0 + 4J_1 + 4J_2 - 6J_1' - 2J_2' - 6J_3' - 6J_4' - 6J_5' - 6J_6'
 - 6J_7' - 2J_8', \nonumber \\
E(2) & = & E_0 + 4J_1 + 4J_2 - 2J_1' - 6J_2' - 6J_3' - 6J_4' - 6J_5' - 6J_6'
 - 2J_7' - 6J_8', \nonumber \\
E(3) & = & E_0 + 6J_1 + 6J_2 - 4J_1' - 6J_2' - 6J_3' - 8J_4' - 8J_5' - 6J_6'
 - 6J_7' - 4J_8', \nonumber \\
E(4) & = & E_0 + 2J_1 + 2J_2 - 4J_1' - 6J_2' - 2J_3' - 4J_4' - 4J_5' - 2J_6'
 - 6J_7' - 4J_8', \nonumber \\
E(5) & = & E_0 + 4J_1 + 4J_2 - 6J_1' - 6J_2' - 6J_3' - 2J_4' - 6J_5' - 2J_6'
 - 6J_7' - 6J_8', \nonumber \\
E(6) & = & E_0 + 4J_1 + 4J_2 - 6J_1' - 6J_2' - 2J_3' - 6J_4' - 2J_5' - 6J_6'
 - 6J_7' - 6J_8', \label{em} \\
E(7) & = & E_0 + 6J_1 + 6J_2 - 8J_1' - 6J_2' - 6J_3' - 4J_4' - 4J_5' - 6J_6'
 - 6J_7' - 8J_8', \nonumber \\
E(8) & = & E_0 - 2J_1 - 2J_2 - 6J_2' - 6J_3' - 6J_6' - 6J_7', \nonumber \\
E(9) & = & E_0 - 2J_1' - 6J_2' - 6J_3' + 2J_4' - 2J_5' - 6J_6' - 2J_7' - 2J_8',
\nonumber \\
E(10) & = & E_0 - 2J_1' - 2J_2' - 6J_3' - 2J_4' + 2J_5' - 6J_6' - 6J_7' - 2J_8',
\nonumber \\
E(11) & = & E_0 - 2J_1 - 2J_2 + 2J_1' - 4J_2' - 4J_3' - 2J_4' - 2J_5' - 4J_6'
 - 4J_7' - 2J_8'. \nonumber 
\end{eqnarray}
\end{widetext}

\twocolumngrid

\begin{figure*}[t]
\includegraphics[width=0.85\linewidth]{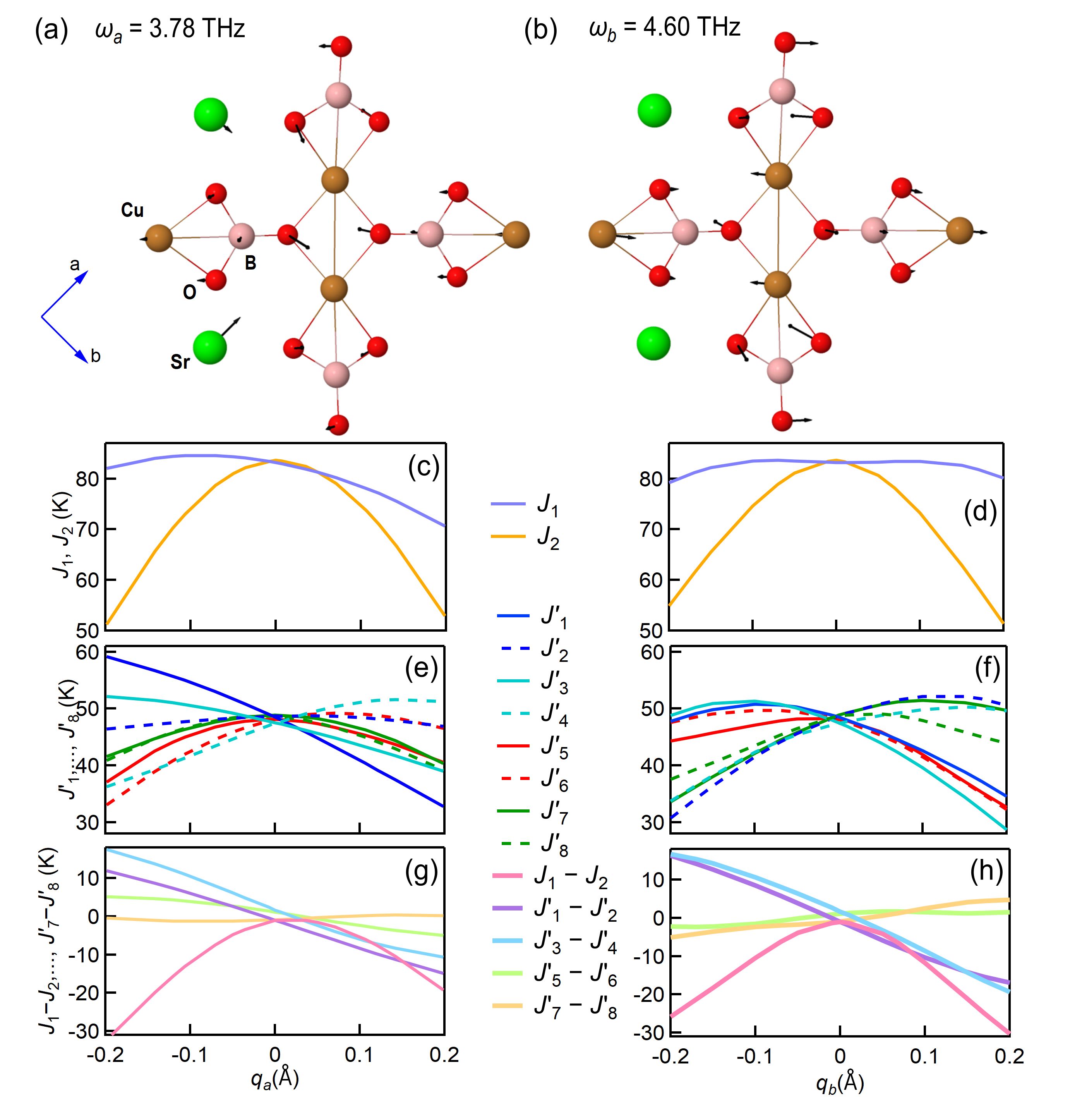}
\caption{{\bf Phonon displacement vectors and calculated magnetic interaction 
parameters in SrCu$_2$(BO$_3$)$_2$.} (a) Atomic motions in the normal mode 
($q_a$) of the lattice at $\omega_a = 3.80$ THz. (b) As in panel (a) for $q_b$ 
at $\omega_b = 4.60$ THz. We comment that these phonon modes are not symmetric 
between the $a'$ and $b'$ axes [Fig.~\ref{uqmf1}(a)], but that their 
doublet counterparts within each $E$-symmetric manifold restore this symmetry. 
(c-d) Intradimer interactions, $J_1$ and $J_2$, shown as functions of the 
phonon displacement amplitudes. (e-f) Interdimer interactions, $J_1'$, \dots 
$J_8'$. (g-h) Differences, $\Delta J_{12}'$, $\Delta J_{34}'$, $\Delta J_{56}'$, 
and $\Delta J_{78}'$, between pairs of interdimer interaction parameters. The 
parameters shown in panels (c-f) were obtained using 11 spin configurations 
and, after verification of their systematic evolution, were centered on the 
results of Table II.}
\label{uqmfs8}
\end{figure*}

\begin{table*}[t]
\caption{\label{t2} Magnetic interaction parameters of SrCu$_2$(BO$_3$)$_2$ 
calculated using the frozen-phonon protocol with all phonon displacements 
set to zero.} 
\centering
\begin{tabular}{c|cccccccccc}
Interaction $\,$ & $J_1$ (K) & $J_2$ (K) & $J_1'$ (K) & $J_2'$ (K) & $J_3'$ (K) 
& $J_4'$ (K) & $J_5'$ (K) & $J_6'$ (K) & $J_7'$ (K) & $J_8'$ (K) \\ \hline
Strength $\,$ & 83.2 & 83.6 & 48.5 & 48.7 & 47.5 & 47.4 & 48.1 & 48.1 & 48.9 
& 48.7
\end{tabular}
\end{table*}

The extraction of the magnetic interaction parameters is by its nature a 
statistical exercise, because different spin configurations lead to different 
local spin densities in the DFT wave function, which cause subtle differences 
in the results for the effective $J$ and $J'$ parameters. We benchmark the 
accuracy of our statistics by testing the magnetic interactions at equilibrium 
(i.e.~$q_a$ and $q_b = 0$) with 100 different spin configurations and performing
a least-squares regression analysis. As we show in Table II, our results are 
fully consistent with the equilibrium $J$ and $J'$ values. The resulting 
statistical error on $J'$ is 0.4 K. $E_0$ in Eqs.~(\ref{em}) is a large 
constant that captures all of the nonmagnetic contributions to the calculation 
and cancels from the equations determining the magnetic interaction parameters.

\subsection{DFT calculations with driven phonons}
 
To model our experiment, in Figs.~\ref{uqmfs8}(a-b) we show the vectors, 
meaning the ensembles of atomic displacements, of the two primary IR-active 
phonons excited by the THz pump ($\omega_a = 3.80$ and $\omega_b = 4.60$ THz). 
We performed DFT calculations of the magnetic interactions in the presence 
of phonon oscillations for each phonon separately, as shown in 
Figs.~\ref{uqmfs8}(c-h), and with both phonons superposed, as we show in the 
time series illustrated in Fig.~\ref{uqmf5}(d). Considering first the 
individual phonons, we observe that the intradimer interactions, $J_1 (q_{a,b})$ 
and $J_2 (q_{a,b})$, have a largely quadratic dependence on $q_{a,b}$ for both 
phonons [Figs.~\ref{uqmfs8}(c-d)], suggesting that out-of-plane O atomic 
motions cause the predominant effects on these parameters. By contrast, most 
of the interdimer interactions, $J_i' (q_{a,b})$, show strong linear as well as 
quadratic contributions [Figs.~\ref{uqmfs8}(e-f)] that depend both on the 
interaction pathway in question and on the combination of in- and out-of-plane 
atomic motions. It is clear that all four difference parameters, $\Delta J_{12}' 
(q_{a,b})$, $\Delta J_{34}' (q_{a,b})$, $\Delta J_{56}' (q_{a,b})$, and $\Delta 
J_{78}' (q_{a,b})$, have strong linear contributions [Figs.~\ref{uqmfs8}(g-h)] 
that allow an efficient driving of two-triplon creation processes by the 
IR-active phonon oscillations (Sec.~IV). 

For the purposes of creating nonlinear magnetophononic phenomena in 
SrCu$_2$(BO$_3$)$_2$, the strongest modulation of the $\Delta J_{i,i+1}' (q_{a,b})$
interactions is produced by the 4.60 THz phonon [Fig.~\ref{uqmfs8}(b)]. This is 
due largely to the fact that its maximum THz-induced phonon displacement, 
$\delta_b = 0.17$ \AA~(Sec.~III), is significantly greater than that of the 
other phonons (at 3.80 THz we estimated the displacement $\delta_a = 0.04$ 
\AA). These maximum displacements are included in the calculations shown in 
Fig.~\ref{uqmf5}, where we have computed the phonon-induced modulation of the 
magnetic interaction parameters as a time series based on the pulse durations 
of our experiment. For this calculation, we computed 285 points covering 10 
ps with a time resolution of 0.035 ps, which accessed both the high- and 
low-frequency regimes to a sufficient degree that we obtained well-resolved 
results for all features, in particular the TBS peak. 

Finally, we comment that an average over all our computed values of $J$ and 
$J'$ indicates that the averaged coupling ratio, $\bar{\alpha} (q_{a,b}) = 
\bar{J}'/\bar{J}$, increases as a function of $q_{a,b}$ for both phonon modes. 
This leads to the results shown in Fig.~\ref{uqmf5}(g) and to the 
possibility of driving the static QPT of the spin system into the plaquette 
phase by using the ultrafast driving of IR phonons to increase the 
time-averaged coupling ratio. 

\end{appendix}

\end{document}